\documentclass[iop,numberedappendix]{emulateapj}
\usepackage{epsfig}
\usepackage{amsmath}
\usepackage{enumerate}
\usepackage{natbib}
\usepackage{hyperref}

\usepackage{ulem}
\usepackage[squaren, Gray, cdot]{SIunits}

\usepackage{color}

\shorttitle{TeV emission from GRBs}
\shortauthors{Vurm et al.}

\def\me{m_{\rm e}}

\def\epsrad{\varepsilon_{\rm rad}}

\def\epsB{\varepsilon_B}

\def\sigmaT{\sigma_{\rm T}}

\def\me{m_{\rm e}}

\def\ba{\begin{align}}
\def\ea{\end{align}}

\def\uX{u_{\rm X}^{\prime}}

\def\E{{\cal E}}
\def\mprot{m_{\rm p}}
\def\epm{e^{\pm}}
\def\Rpm{R_{\rm load}}

\def\dL{d_{\rm L}}

\def\Rdec{R_{\rm dec}}
\def\RdecW{R_{\rm dec}^{\rm wind}}
\def\RdecI{R_{\rm dec}^{\rm ISM}}
\def\tdecI{t_{\rm dec}^{\rm ISM}}

\def\tendI{t^{\rm ISM}_{\rm 100 GeV}}
\def\tendW{t^{\rm wind}_{\rm 100 GeV}}

\def\tauWload{\tau_{\gamma\gamma, {\rm load}}^{\rm wind}}
\def\tauIdec{\tau_{\gamma\gamma, {\rm dec}}^{\rm ISM}}

\def\tdec{t_{\rm dec}}
\def\tGRB{t_{\rm GRB}}
\def\Gjet{\Gamma_{\rm jet}}
\def\LGRB{L_{\rm GRB}}
\def\EGRB{{\cal E}_{\rm GRB}}
\def\Ekin{{\cal E}_{\rm kin}}
\def\Lkin{L_{\rm kin}}
\def\Emax{E_{\rm max}}
\def\EX{{\cal E}_{\rm X}}
\def\gth{\gamma_{\rm inj}}
\def\epse{\varepsilon_{\rm e}}
\def\epsX{\varepsilon_{\rm X}}
\def\DX{\Delta_{\rm X}}
\def\mue{\mu_{\rm e}}
\def\tpm{t_{\rm load}}
\def\A{A_{\rm eff}}
\def\epsTeV{\varepsilon_{\rm TeV}}
\def\taugg{\tau_{\gamma\gamma}}

\def\tgg{t_{\gamma\gamma}}
\def\lX{l_{\rm X}}
\def\LX{L_{\rm X}}
\def\Zpm{Z_{\pm}}

\def\tIC{t_{\rm IC}^{\prime}}
\def\tsyn{t_{\rm syn}^{\prime}}
\def\tdyn{t_{\rm dyn}^{\prime}}

\def\FlTeV{{\cal F}_{\rm TeV}}
\def\FlGRB{{\cal F}_{\rm GRB}}

\def\Gfs{\Gamma_{\rm fs}}
\def\Grs{\Gamma_{\rm rs}}
\def\tfs{t_{\rm fs}}

\def\Rrs{R_{\rm rs}}
\def\bjet{\beta_{\rm jet}}
\def\brs{\beta_{\rm rs}}

\def\Swift{{\it Swift }}
\def\Fermi{{\it Fermi }}

\def\Eq{Equation}

\def\xipm{\xi_{\rm load}}

\newbox\grsign \setbox\grsign=\hbox{$>$} \newdimen\grdimen \grdimen=\ht\grsign
\newbox\simlessbox \newbox\simgreatbox \newbox\simpropbox
\setbox\simgreatbox=\hbox{\raise.5ex\hbox{$>$}\llap
     {\lower.5ex\hbox{$\sim$}}}\ht1=\grdimen\dp1=0pt
\setbox\simlessbox=\hbox{\raise.5ex\hbox{$<$}\llap
     {\lower.5ex\hbox{$\sim$}}}\ht2=\grdimen\dp2=0pt
\setbox\simpropbox=\hbox{\raise.5ex\hbox{$\propto$}\llap
     {\lower.5ex\hbox{$\sim$}}}\ht2=\grdimen\dp2=0pt
\def\simgt{\mathrel{\copy\simgreatbox}}
\def\simlt{\mathrel{\copy\simlessbox}}

\begin{document}

\title{On the prospects of gamma-ray burst detection in the TeV band}

\author{
Indrek Vurm and Andrei M. Beloborodov
}
\affil{$^1$Physics Department and Columbia Astrophysics Laboratory, Columbia University, 538 West 120th Street, New York, NY 10027, USA \\
}

\label{firstpage}
\begin{abstract}
A gamma-ray burst (GRB) jet running into an external medium is expected to generate luminous GeV-TeV emission lasting from minutes to several hours. The high-energy emission results from inverse Compton upscattering of prompt and afterglow photons by shock-heated {\it thermal} plasma. At its peak the high-energy radiation carries a significant fraction of the power dissipated at the forward shock. We discuss in detail the expected TeV luminosity, using a robust ``minimal'' emission model. Then, using the statistical properties of the GRB population (luminosity function, redshift distribution, afterglow energy) we simulate the expected detection rates of GRBs by current and upcoming atmospheric Cherenkov instruments. We find that GRBs exploding into a low-density interstellar medium must produce TeV emission that would have already been detected by the currently operating Cherenkov telescopes. The absence of detections is consistent with explosions into a dense wind of the GRB progenitor. If the typical environment of long GRBs is a Wolf-Rayet progenitor wind, as suggested by the recent analysis of Fermi LAT data, then the upcoming Cherenkov Telescope Array (CTA) should be able to detect a fraction of GRBs that trigger the space-borne detectors. Since absorption by the extragalactic background light limits the detectability above 0.1~TeV for all but the most nearby bursts ($z\lesssim 1$), the reduced energy threshold of CTA is the key improvement over current instruments, which should boost the detection rate by at least a factor of 3.
\end{abstract}

\bigskip


\bigskip


\section{Introduction}

The Compton Gamma Ray Observatory and the \Fermi telescope have established
gamma-ray bursts as luminous GeV sources.
To date, the LAT instrument onboard \Fermi has detected high-energy emission 
($> 20$~MeV) from over 100 GRBs.
The peculiar and recurrent properties of the GeV light curves
have allowed significant recent progress in the understanding of
the external blast-wave properties at early stages
as well as the environments they explode into. The early GeV emission arises from
inverse Compton emission (IC) from the forward shock 
created by a blast wave running into a pair-loaded external medium
(\citealt{BHV14}, hereafter B14; \citealt{Vurm14,Hascoet15}).
Detailed modeling of radiative transfer and blast wave dynamics has allowed 
one to use the GeV data to unambiguously 
reconstruct the explosion parameters such as the kinetic energy, initial Lorentz factor etc. 
in a handful of bright GRBs, as well as determine the density and profile of the ambient 
medium. These models also predict that a significant fraction of the IC energy is emitted 
in the TeV band, at a level that in several cases could be detectable by ground-based 
Cherenkov observatories. Quantifying the last statement for the entire GRB population 
constitutes the main focus of the present paper.

The idea that GRBs could be luminous in the GeV-TeV range was proposed more than 
two decades ago. \citet{Meszaros1994} suggested that the inverse Compton scattering
could be the dominant cooling mechanism of shock-heated/accelerated electrons in the 
external blast wave close to the deceleration radius (where the external shock luminosity 
peaks), and give rise to multi-GeV emission that is delayed with respect to the main MeV 
component. If the ambient medium is the dense progenitor wind,
the transition to the deceleration regime typically occurs at a time comparable to the 
prompt duration (i.e. the reverse shock is at least mildly relativistic).
However, one should keep in mind that during the first minute after the explosion,
the prompt radiation field ahead of the forward shock
interacts strongly with the ambient medium and loads it with copious electron-positron pairs
(\citealt{ThompsonMadau00,B02a}; B14). Consequently, the energy {\it per particle} at 
the shock is reduced and both GeV and TeV luminosities are suppressed.
As the blast wave expands, the pair loading decreases and the characteristic IC energy 
of the shocked electrons evolves first into the GeV and then into the TeV band.
This evolution is responsible for the delay of the (observed) GeV onset with respect to 
the prompt emission, and predicts an additional delay of the TeV radiation (B14).
The TeV radiation can be further delayed by $\gamma\gamma$-opacity effects
\citep[see e.g.][]{Baring2006}. Thus neither the GeV nor TeV peak is necessarily
associated with the start of the self-similar deceleration regime.

Among different possibilities, inverse Compton scattering has 
been established as the main candidate for the high-energy emission.
The main alternative discussed in the literature, synchrotron emission,
in unable to generate photons above a few GeV, particularly at late times
\citep[see e.g.][]{PiranNakar10}.
Mechanisms invoking hadronic processes suffer from poor efficiency
\citep[e.g.][]{Asano2009,Asano2012}. A systematic study of possible
radiation mechanisms for the multiwavelength afterglow emission (electron IC, electron 
and proton synchrotron) was conducted by \citet{Zhang2001}.
For a broad range of parameters IC emission
was found to be the most favorable for GeV-TeV production.
Perhaps the strongest evidence to date in favor of the IC mechanism
comes from the successful fitting of the entire GeV light curve for seven LAT burst using
a detailed IC radiative transfer model with only three adjustable parameters
\citep{Hascoet15}. When a magnetic field is added as a parameter to describe the 
synchrotron counterpart of the GeV flash (detected in two GRBs), the optical light curve is 
remarkably well reproduced by the same model.
In view of all of these results, we will consider only the IC mechanism as the source of
very high-energy (VHE) photons in this paper.

The electrons energized by the forward shock are fast-cooling and radiatively efficient 
for up to several hours (in some cases days) after the explosion;
most of their energy is radiated as IC emission as long as
the electron energy fraction at the forward shock exceeds that of the $B$-field, 
$\epse/\epsB>1$ \citep[e.g.][]{Sari2001}.
In many cases the (theoretical) IC spectrum extends well into the VHE band
(B14; \citealt{Hascoet15}; see also \citealt{Sari2001,Fan2008}).
where the emission can last for up to a few hours.
It is important to note that long-lasting VHE emission does not rely on efficient electron 
{\it acceleration} at the shock;
it is sufficient that the shocked baryons share a non-negligible fraction of their energy with
{\it thermal} electrons via collective electromagnetic interactions.

So far there have been no definitive detections of GRBs by Cherenkov observatories,
while several groups have reported upper limits for a number of bursts
\citep[e.g.][]{Horan2007,Albert2007,Aharonian2009,Jarvis2010}.
Tentative low-level detections ($<3\sigma$) have been reported by \citet{Atkins2000}
and \citet{Poirier2003} in GRB 970417a and GRB 971110, respectively.
While there have been a number of GRBs with good photon statistics in the LAT band,
a significant turnover/cutoff in the high-energy spectrum
has not been detected in all but one (GRB 090926A) bursts.
This suggests that a significant fraction of of LAT bursts
could have spectra extending beyond $\sim 100$~GeV,
which would make them promising targets for ground-based detection.
Compared to space-borne experiments such as Fermi/LAT, ground-based Cherenkov 
observatories benefit from $4$-$5$ orders of magnitude larger effective area.
On the other hand, EBL absorption limits detectability above a few $100$~GeV for all but the 
nearest GRBs, which is near the low energy threshold of current instruments.
Therefore, efforts to lower the threshold as well as improve sensitivity at 
$\lesssim 100$~GeV in next generation instruments such as the Cherenkov Telescope 
Array (CTA) hold the key for routine detection of GRBs in the VHE band.

In this work our goal is to employ a simple IC model for TeV emission with minimal assumptions
to assess the detection prospects for ground-based Cherenkov observatories,
using a Monte Carlo simulation based on observational data on the luminosity function,
redshift distribution and statistical properties of X-ray afterglows.
In our "minimal" model we are concerned only with high-energy emission from thermal particles at the forward shock, which are robustly predicted by 
{\it ab initio} plasma simulations of relativistic collisionless shocks
\citep[e.g.][]{Spitkovsky2008a,SironiSpitkovsky2011}.
The details of an additional population of accelerated {\it non}-thermal particles are on a somewhat less firm theoretical footing,
and depend more sensitively on the conditions at the shock, in particular the strength and configuration of the upstream magnetic field.
Even though accelerated particles must be present to generate the broad-band synchrotron afterglow commonly observed in GRBs, their 
energy budget is below that of the thermal component, and they do not dominate 
the observed high-energy emission (B14). Similar to B14, we will avoid introducing more 
parameters to describe the additional emission from nonthermal electrons and focus on 
the emission from the thermal plasma. This defines our ``minimal'' model for TeV emission
from a GRB blast wave.

The emission mechanism can be outlined as follows.
The forward shock of the explosion runs into the external medium, which is heavily 
loaded with $e^\pm$ pairs at small radii and becomes pair-free at radii $R>\Rpm$
(calculated directly from the observed luminosity and spectrum of the prompt 
GRB emission, see B14). In the shock rest frame, most of the kinetic energy of the 
upstream material is carried by baryons.
Typically a few tens of percent of this energy is communicated to leptons at the shock,
creating relativistically hot $e^\pm$ plasma behind the shock.
The heated leptons rapidly cool by upscattering  
the prompt GRB radiation (as long as it overlaps with the blast wave) 
and later by upscattering synchrotron
afterglow radiation emitted by the blast wave itself.
The TeV production starts when the pair loading has dropped sufficiently low so that
the characteristic IC photon energy of thermal leptons has evolved into the TeV band.
The maximum photon energy is attained 
at $R\approx\Rpm$ and typically ranges from a few hundred GeV to a few TeV.

In many bursts the VHE radiation is initially absorbed by
the X-ray afterglow photons and reprocessed to lower (GeV) energies.
The observed onset is then delayed until the source becomes optically thin to TeV photons.
As the blast wave decelerates, the energy of IC photons from thermal electrons
eventually falls below the TeV band and VHE emission  is expected to decline,
typically a few hours after the trigger. Overall, the rapid IC cooling of the electrons
allows a significant fraction of the shock-dissipated power to be radiated at high energies.
This emission only remains hidden if
the IC photon energy falls below the TeV band before the source becomes transparent to those photons, which is the case for weak bursts exploding into a dense progenitor wind.

The paper is organized as follows. In Section \ref{sec:TEVemiss} we discuss the
mechanism of VHE emission from GRB blast waves.
A brief discussion of the sensitivity of current and upcoming Cherenkov instruments 
and an estimate for the expected photon counts is given in Section \ref{sec:det}.
Section \ref{sec:MC} describes the Monte-Carlo procedure for simulating 
a large population of GRBs and their detection by the Cherenkov telescopes. 
The results of our numerical modeling are presented in Section \ref{sec:res},
and our main conclusions are summarized in Section \ref{sec:concl}.


\section{Expected TeV emission}

\label{sec:TEVemiss}

\subsection{Conditions for bright TeV emission}

The key properties of the external blast wave that lead to significant emission
at $\gtrsim 0.1$~TeV during the first several minutes can be summarized as follows:
\begin{enumerate}
\item
The thermal electrons behind the forward shock receive a significant fraction of the 
energy dissipated at the shock;
a typical value found in PIC simulations of collisionless shocks is $\epse \approx 0.3$
\citep{SironiSpitkovsky2011}, which we will adopt here.
\item
The blast wave is radiatively efficient, i.e.
the (thermal) leptons are in the fast cooling regime at the early stages of the afterglow,
and radiate most of their energy by the IC scattering of the prompt and afterglow radiation.
\item
The characteristic energy of the shocked thermal electrons is
$\gth\approx \Gamma \mue \epse \mprot/(\me \Zpm)$,
where $\Zpm$ is the pair loading factor (number of $\epm$ pairs per proton),
and $\mue=2$ is the ion mass per proton in the wind medium.
The corresponding maximal IC energy is
\begin{align}
E_{\rm max} = \Gamma\gth\me c^2 \approx 5 \, \Zpm^{-1} \left(\frac{\mue}{2}\right) \left(\frac{\epse}{0.3}\right) \left(\frac{\Gamma}{100}\right)^2 \, \mbox{TeV},
\label{eq:EIC}
\end{align}
which remains above $\sim 0.1$~TeV as long as $\Gamma\gtrsim 15$ (when $\Zpm\approx 1$, i.e. $R>\Rpm$).
The thermal TeV emission thus ceases 
only after the blast wave has undergone significant deceleration;
it can last up to several hours.
\end{enumerate}

The typical GRB Lorentz factors $\Gjet = 100-1000$ are more than sufficient
to generate TeV photons at the forward shock even without
any non-thermal particle acceleration.
However, early TeV emission is suppressed due to two effects:
\begin{enumerate}
\item
Opacity due to $\gamma\gamma$ pair production.
Both the prompt and afterglow radiation can contribute to 
$\gamma\gamma$ opacity.
The prompt radiation is emitted at much smaller radii, however it dominates the 
$\gamma\gamma$ opacity until it has completely overtaken the relativistic blast wave.
The prompt radiation
is almost perfectly beamed at radii where 
the bulk of the TeV radiation is emitted.
Therefore, there exists an ``escape cone'', i.e. photons with sufficiently small angles with respect to the radial direction can 
escape (see e.g. B14).
Even when a typical optical depth created by the prompt radiation for the TeV photons is 
large, $\taugg\gg 1$, a fraction $\sim \taugg^{-1}$ of TeV photons will escape in the cone.
By contrast, the X-ray afterglow radiation is roughly isotropic in the blast-wave frame,
and therefore exponentially suppresses the high-energy emission.
The transition to $\gamma\gamma$ transparency
is discussed in more detail in Section \ref{sec:gammagamma}.

\item
Reduction of $E_{\rm max}\propto \Zpm^{-1}$ below the TeV band due to pair 
loading of the external medium (\Eq~\ref{eq:EIC}).
The prompt radiation ahead of the forward shock pre-loads the external medium with 
$Z_\pm>1$ pairs per proton at radii $R<\Rpm$ \citep[see][]{B02a},
\begin{align}
\Rpm &= \left(\frac{\sigmaT\EGRB}{4\pi\me c^2\xi}\right)^{1/2}	\nonumber \\
&\approx 1.8\times 10^{16} \, \E_{{\rm GRB}, 53}^{1/2} \left(\frac{\xipm}{20}\right)^{-1/2} \,\, \mathrm{cm},
\label{eq:Rpm}
\end{align}
where $\E_{\rm GRB}$ is the energy of the prompt radiation ahead of the forward shock,
normalized to $10^{53}$~erg. Here $\xipm$ is the characteristic
column density of the prompt radiation ahead of the forward shock 
that triggers exponential runaway of $e^\pm$ loading; its value
depends only on the spectral shape of the prompt radiation  (but {\it not} its luminosity)
and typically varies between $10-30$.
The TeV emission is delayed until $E_{\rm max}$ evolves into the VHE band;
$E_{\rm max}$ attains its maximal value at $R \approx \Rpm$.
\end{enumerate}

\subsection{Wind medium vs. ISM}

\label{sec:windISM}

GRBs with massive progenitors of Wolf-Rayet type are expected to explode into a 
dense progenitor wind with mass density $\rho\propto R^{-2}$, made of elements 
heavier than hydrogen. Strong evidence for the wind external medium is provided by the 
reconstruction of early GeV emission observed by {\it Fermi} LAT and its optical counterpart 
\citep{Hascoet15}. We will also consider an alternative scenario where the explosion 
occurs in an approximately uniform interstellar medium (ISM).
The TeV emission predicted by the two models is quite 
different, and hence TeV observations can provide an additional test to distinguish
between the two possibilities.
The key difference is the enhanced density at $R\lesssim 10^{17}$~cm in the wind case,
which results in earlier blast wave deceleration. Among other effects discussed below,
this leads to a higher comoving radiation density
and therefore a higher intrinsic opacity to TeV photons during the time of their production.
Below we first consider the ISM and then the wind medium.

In the ISM the deceleration radius is
\begin{align}
\RdecI =
\left(
\frac{3\Ekin}{8\pi\mprot c^2 n \Gjet^2}
\right)^{1/3}=
9.3\times 10^{16} \,\, \frac{\E_{{\rm kin}, 53}^{1/3}}{n^{1/3} \, \Gamma_{{\rm jet}, 2}^{2/3}} \,\, \mbox{cm},
\label{eq:app:Rdec}
\end{align}
where 
$\Ekin$ is the kinetic energy of the ejecta (isotropic equivalent), and 
$n\sim 1$~cm$^{-3}$ is the typical number density of the ISM.
The deceleration radius is typically larger than $\Rpm$, and hence
the jet will coast with $\Gamma\approx \Gjet$ between $\Rpm$ and $\RdecI$ while
$E_{\rm IC,max}$ stays at its maximum.
The maximal thermal IC photon energy
is well above $100$~GeV and
can reach $\gtrsim 100$~TeV in fast jets
according to Equation (\ref{eq:EIC}) (where now $\mue=1$, 
as the ISM mainly consists of hydrogen). Since $\Rpm<\RdecI$,
only a fraction of the blast-wave energy is dissipated near $\Rpm$,
and its luminosity continues to rise as the blast wave expands to $\RdecI$. 
The peak luminosity is achieved at $R\sim\RdecI$, 
which corresponds to observer time
\begin{align}
\tdecI \approx \frac{\RdecI}{c \Gjet^2} = 310 \,\, \frac{\E_{{\rm kin}, 53}^{1/3}}{n^{1/3} \, \Gamma_{{\rm jet}, 2}^{8/3}} \,\, \mbox{s}.
\label{eq:tdec:ISM}
\end{align}
It will be shown below that the TeV emitting electrons are radiatively efficient at $\RdecI$ in typical bursts,
and the source is transparent to $0.1-1$~TeV gamma rays.
The TeV emission is therefore expected to peak at $t\approx \tdecI$.
The VHE emission from the thermal shock plasma
ceases when the jet has decelerated to $\Gamma\approx 20$, which occurs at
\begin{align}
\tendI = 6.6\times 10^3 \,
\frac{\E_{{\rm kin},53}^{1/3}}{n^{1/3}} \, \left(\frac{\epse}{0.3} \right)^{4/3} \,\,\mathrm{s},
\label{eq:tend:ISM}
\end{align}
where we have used 
Equation (\ref{eq:EIC}) with $\Zpm=1$, 
and $\Gamma \propto t^{-3/8}$ appropriate for self-similar deceleration in a constant density external medium.

Next consider a blast wave in the wind medium. If the reverse shock is non-relativistic,
the deceleration radius is given by
\begin{align}
\label{eq:RdecW}
\RdecW = \frac{\Ekin}{8\pi c^2 A \Gjet^2} = 4.4\times 10^{15} \, \frac{\E_{{\rm kin},53} }{ A_{\rm 11} \, \Gamma_{{\rm jet}, 2}^{2} } \,\, \mathrm{cm},
\end{align}
where 
$A=\rho R^2 = \mbox{constant}$ is the wind density parameter, which we normalize to 
$10^{11}$~g~cm$^{-2}$, a characteristic value expected for a Wolf-Rayet wind.\footnote{If 
    $\RdecW\ll\Rpm$, \Eq~(\ref{eq:RdecW}) somewhat underestimates $\RdecW$, as
    the blast-wave deceleration is delayed due to pre-acceleration of the external medium 
    by the prompt radiation
    \citep{B02a}.}
For typical GRB parameters, however, the reverse shock is relativistic 
\citep[e.g.][]{Chevalier1999}.
The self-similar deceleration of the ejecta begins when the reverse shock has crossed the 
ejecta, and the corresponding observed time approximately equals the duration of the ejecta,
which is comparable with the observed GRB duration $\tGRB$. Thus, a simple estimate 
for the deceleration time is given by $\tdec\sim\tGRB$.

The deceleration time is typically shorter than the time it takes the blast wave to
reach $\Rpm$. Then using the self-similar deceleration solution for an adiabatic blast wave,
\begin{align}
\Gamma(\tpm) &= 
\left. 
\left(\frac{\Ekin}{8\pi c^2 A R}
\right)^{1/2}
\right|_{\tpm}			\nonumber \\
&= 50 \,\,  \frac{ \E_{{\rm kin},53}^{1/2} }{ A_{11}^{1/2} \E_{{\rm GRB},53}^{1/4}  } \left(\frac{\xipm}{20}\right)^{1/4},
\label{eq:Gpm}
\end{align}
we estimate that $Z_\pm$ drops to unity at the observed time
\begin{align}
\tpm \approx \left. \frac{R}{2c\Gamma^2}\right|_{\tpm}
= 120
\,  \frac{A_{11}}{\E_{{\rm kin},53}} \, \E_{{\rm GRB},53} \left(\frac{\xipm}{20}\right)^{-1}	\,\, \mathrm{s}.
\label{eq:tpm}
\end{align}
The IC photon energy reaches the maximum at this time, shaped by two competing 
effects: the exponential reduction of $Z_\pm$ and the power-law decline of $\Gamma$.
The maximal IC photon energy is found from Equations (\ref{eq:EIC}) and (\ref{eq:Gpm}),
\begin{align}
\Emax(\tpm) &\approx
\left. \Gamma\gth \me c^2 \right|_{\tpm} \approx
\left.\Gamma^2 \, \mue \epse \mprot c^2 \right|_{\tpm}							\nonumber \\
&= 1.4 \, \frac{\E_{{\rm kin},53}}{A_{11} \, \E_{{\rm GRB}, 53}^{1/2}}  \, \left(\frac{\epse}{0.3} \right) \,\, \mathrm{TeV}.
\label{eq:Emaxmax}
\end{align}
After $\tpm$, the maximal photon energy decreases as $\Emax = \Emax(\tpm)(t/\tpm)^{-1/2}$
due to the decrease in $\Gamma$; $\Gamma$ reaches $20$ and $E_{\rm IC}$
falls below $100$~GeV at
\begin{align}
\tendW
= 2.3\times 10^{4} \,  \frac{\E_{{\rm kin},53}}{A_{11}} \, \left(\frac{\epse}{0.3} \right)^2 \,\,\mathrm{s}.
\label{eq:t100}
\end{align}
One can see from Equations (\ref{eq:tpm}), (\ref{eq:Emaxmax}) and (\ref{eq:t100}) that
the time when the spectrum extends to the highest energy, the maximal IC photon energy,
as well as the time when the TeV emission ends are
controlled by essentially a single unknown parameter $\Ekin/A$.
Note also that in most energetic bursts ($\E_{{\rm kin}} \gtrsim 10^{54}$~erg)
emission above $100$~GeV can last for more than a day.

\subsection{Blast wave luminosity}
\label{sec:TeVlum}

An upper limit on the expected TeV emission is set by the total power dissipated in 
the blast wave.
The dissipated luminosity at the forward shock
in the self-similar deceleration regime is
$L_{\rm diss} \approx \chi\Ekin/t$,
where $\chi = 1/4$ or $3/8$ in the wind and ISM cases, respectively.
The energy given to
thermal electrons per logarithmic time interval is
\begin{align}
t L
\approx \chi\epse\,\Ekin \approx 10^{52} \, \E_{{\rm kin},53} \, \left(\frac{\epse}{0.3} \right) \left(\frac{\chi}{0.3} \right) \,\, \mathrm{erg}.
\label{eq:tLIC}
\end{align}
Strictly speaking, $\Ekin$ in the above equation should refer to the blast wave energy 
{\it at time} $t$;
$\Ekin = \mbox{const}$ only in an adiabatic blast wave.
We will show below that at times of interest the electrons at the forward shock are fast cooling,
thus a fraction $\epse\approx 0.3$ of the energy dissipated at the forward shock is lost to radiation.
The total energy of the blast wave then decreases approximately as $\Ekin \propto \Gamma^{\epse}$;
the blast wave loses approximately half of its energy
by the time $\Gamma\approx 20$, i.e. when the TeV emission ceases.
This effect was ignored in the above analytic estimates, however it will be taken 
into account in our numerical simulations presented below.

The reduction of $\Ekin$ due to radiative
losses can be even more severe at early stages
when the pair loading is high, $\Zpm \gtrsim \mue\mprot/\me \sim 3000$,
before the onset of GeV/TeV emission (B14).
Then the electron/positron pairs (rather than baryons)
carry most of the inertia of the medium ahead of the forward shock.
The energy dissipated at the shock is given directly to the
pairs, resulting in a radiative blast wave that loses nearly
$100\%$ of the dissipated energy to radiation.
In the ISM case this effect is unlikely to be important as the pairs disappear well before
the blast wave can communicate a significant fraction of its energy to the external medium ($\Rpm<\RdecI$).
In the dense progenitor wind, however, the losses can be significant.
A more detailed discussion of the radiative phase is given in Section \ref{sec:radloss} below.

The fraction of energy given by Equation (\ref{eq:tLIC}) that emerges
above $100$~GeV depends on:
(1) whether the electrons are fast-cooling, i.e. efficiently radiate their energy,
(2) attenuation due to $\gamma\gamma$- absorption on afterglow radiation,
(3) the fraction of energy lost to low-energy synchrotron emission,
(4) the inverse Compton spectrum, which in turn depends on the slope of the X-ray 
afterglow spectrum that constitutes the target radiation for IC scattering.
Sections~\ref{sec:gammagamma}-\ref{sec:syn} below discuss each of these factors in detail.

\subsection{$\gamma\gamma$-absorption}
\label{sec:gammagamma}

In contrast to GeV emission observed by LAT, the
bulk of TeV radiation in a typical burst is 
expected to emerge after the end of the prompt emission. Then
the prompt radiation has already decoupled from the ejecta,
so the target photons for IC scattering must be provided by the afterglow radiation.
High $\Gamma$ bursts exploding into the low-density ISM may be an exception to this expectation,
however both current and next generation Cherenkov observatories are 
anyway unlikely to catch a significant number of GRBs in the prompt phase.
For these reasons we will concentrate on 
afterglow as a source of targets for both IC scattering and $\gamma\gamma$ absorption.

The main parameter that determines the opacity of the source to photon-photon absorption is 
the compactness of the X-ray afterglow radiation field, defined as
\begin{align}
\lX &= 
\frac{\sigmaT}{\me c^2} \uX \frac{R}{\Gamma}	
= \frac{\sigmaT}{4\pi \me c^3} \frac{L_{\rm X}}{R \, \Gamma^3}	\nonumber \\
&= 2\times 10^{-4} \, L_{{\rm X}, 48}\, R_{16}^{-1} \, \left(\frac{\Gamma}{100}\right)^{-3} \nonumber \\
&= 
\left\{ \begin{array}{ll}
	8.0\times 10^{-3} \, \frac{ \displaystyle \varepsilon_{{\rm X},-1} A_{11}^{1/2} \, \E_{{\rm GRB}, 53}^{1/4} }{ \displaystyle \E_{{\rm kin},53}^{1/2} \Delta_{{\rm X}, 1}}  , 		& \, \mbox{(wind)} \\ 	
	7.5\times 10^{-5} \, \frac{\displaystyle \varepsilon_{{\rm X},-1} n^{2/3} \E_{{\rm GRB}, 53} \Gamma_{{\rm jet},2}^{1/3}}{\displaystyle \E_{{\rm kin},53}^{2/3} \Delta_{{\rm X}, 1}}	& \, \mbox{(ISM),} 	
	\end{array}
	\right.
\label{eq:lX}
\end{align}
where $L_{\rm X}$ and $\uX$ are the X-ray luminosity and comoving energy density, 
respectively. The characteristic radius is chosen as $\Rpm$ for wind and $\RdecI$ for 
ISM, according to where most of the TeV emission is produced in the two cases. 
In the last line of \Eq~(\ref{eq:lX}) we have parametrized the X-ray luminosity as
$\LX = \epsX\E_{\rm GRB}/(t \DX)$, where $\DX$
is the ``typical" duration of the observed X-ray afterglow in logarithmic units;
we use $\Delta_{\rm X} = 10$ as the fiducial value, which approximately corresponds to 
4 decades in time. Using the empirical relation between the prompt and X-ray afterglow 
energy reported by \citet{Margutti13} in {\it Swift} afterglows, one finds 
$\epsX \approx 0.1 - 0.2$  for $\E_{\rm GRB} = 10^{52}-10^{54}$~erg.

Defining $x = E/(m_e c^2)$ and $x_{\rm t} = E_{\rm t}/(m_e c^2)$ as the dimensionless energies of the primary and target photons,
respectively, one can express the pair-production opacity for the high-energy photon as%
\footnote{Here $\lX$ should refer to
the compactness per logarithmic photon energy interval at the pair production threshold,
rather than that integrated over the bandpass of the X-ray instrument.
If $\beta \approx 1$, 
the difference depends logarithmically on the width of the band and is ignored in the analytical estimates provided here.
}
\begin{align}
\taugg = 
\eta(\beta) \, \frac{\lX}{x_{\rm t}^\prime} \approx
\eta(\beta) \, \lX \, \frac{x}{\Gamma},
\label{eq:opac:gen}
\end{align}
where
the prime refers to the comoving frame of the blast wave.
In this formulation, the target photon energy has to be taken at the threshold 
$x_{\rm t}^\prime = 1/x^{\prime} \approx \Gamma/x$.
The numerical factor $\eta(\beta)$ is defined in \citet{Sve87},
where $\beta$ is the X-ray spectral index;	in the analytical estimates below we will use $\eta(\beta)=0.12$, approproate for $\beta\approx 1.0$ in typical X-ray afterglows
\citep[e.g.][]{Grupe2013,Racusin2016}.

The average energy of the target photons for $\gamma\gamma$ absorption
in an isotropic comoving radiation field is $ x_{\rm t}^\prime \approx 2/x^{\prime}$.
In the lab frame this gives
\begin{equation}
  E_{\rm t} \approx 52 \,
  \left( \frac{\Gamma}{100} \right)^{2} \, \left( \frac{E_{\rm obs}}{100~\mathrm{GeV}} \right)^{-1} \,\,    
  \mathrm{keV}.
\label{eq:Esobs}
\end{equation}
For blast waves in ISM,
the target photons are typically in the hard X-ray band at $\tdecI$ when most of the VHE emission is being emitted.
The pair-production opacity for high-energy photons generated at $\tdecI$ in the ISM is
\begin{align}
\tauIdec =
1.8 \times 10^{-2} \,\,
\frac{\varepsilon_{{\rm X},-1} \, n^{2/3} \,\E_{{\rm GRB}, 53}}{\E_{{\rm kin}, 53}^{2/3} \, \Gamma_2^{2/3} \Delta_{{\rm X}, 1}} \left( \frac{E_{\rm obs}}{100~\mathrm{GeV}} \right).
\label{eq:app:tgg}
\end{align}
Thus radiation up to several TeV can escape the source without significant attenuation.

The wind case is somewhat more complex. 
The typical energy of target photons at the peak of TeV emission is given by
\begin{equation}
  E_{\rm t}^{\rm wind}\stackrel{\tpm}{=} 
12.8 \, 
\frac{\E_{{\rm kin},53}}{A_{11} \, \E_{{\rm GRB}, 53}^{1/2}} \, \left( \frac{E_{\rm obs}}{100~\mathrm{GeV}} \right)^{-1} \,\, \mathrm{keV}.
\end{equation}
The blast wave has significantly decelerated by the time the TeV emission turns on near 
$\Rpm$; accordingly, the target photons have lower energies compared with the ISM case,
and usually fall into the XRT (soft X-ray) band.
The $\gamma\gamma$ optical depth at the pair loading radius is
\begin{align}
\tauWload =
3.8 \,   \frac{\varepsilon_{{\rm X},-1} A_{11} \, \E_{{\rm GRB}, 53}^{1/2}}{\E_{{\rm kin}, 53} \, \Delta_{{\rm X}, 1}} \left( \frac{E_{\rm obs}}{100~\mathrm{GeV}} \right).
\label{eq:taugg}
\end{align}
The optical depth is highest for bursts with a modest energy budget 
that explode into high-density winds and have relatively luminous X-ray afterglows.
Conversely, powerful bursts with weak afterglows
could be transparent to TeV gamma-rays from their onset near $\tpm$.
This was likely the case in the famous and nearby GRB~130427A, which had
$\E_{\rm GRB}\sim 10^{54}$~erg~s$^{-1}$,
$\E_{\rm kin}\sim \mbox{a few}\times 10^{53}$~erg, and
$\epsX\approx \mbox{a few}\times 10^{-2}$ \citep{Golenetskii13,Perley13}.
Note also that  evidence for a wind medium of moderate density
$A\approx 5\times 10^{10}$~g~cm$^{-1}$ was provided for this burst
by the detailed modeling of the GeV+optical flash \citep{Vurm14}.

After $\tpm$, the optical depth
scales as $\taugg(t) = \tauWload (t/\tpm)^{-1/2}$,
and the source becomes optically thin at observer time
\begin{align}
\tgg = 1.8 \times 10^3 \,  \, \frac{\varepsilon_{{\rm X},-1}^2 \, A_{11}^3 \, \E_{{\rm GRB}, 53}^2 }{\E_{{\rm kin}, 53}^3 \, \Delta_{{\rm X}, 1}^2} \, \left( \frac{E_{\rm obs}}{100~\mathrm{GeV}} \right)^2 \, \mathrm{s}.
\label{eq:tgg}
\end{align}
The majority of bursts exploding into the wind medium
are expected to be opaque to $>100$~GeV radiation at $\tpm$.
A high TeV luminosity is expected if $\tgg<t_{\rm 100 GeV}$ (\Eq~\ref{eq:t100}), i.e. the 
blast wave is still producing TeV photons at the transition to $\gamma\gamma$ transparency.
This yields the condition,
\begin{align}
\frac{\E_{{\rm kin},53}}{A_{11}\, \E_{{\rm GRB}, 53}^{1/2}} \, > 0.53 \, 
\frac{ \varepsilon_{{\rm X},-1}^{1/2} }{ \Delta_{{\rm X}, 1}^{1/2} } \, 
\left( \frac{E_{\rm obs}}{100~\mathrm{GeV}} \right) \,\left(\frac{\epse}{0.3} \right)^{-1/2}.
\label{eq:yes}
\end{align}
If this condition is satisfied and $\tgg>\tpm$
then the TeV emission peaks at $\tgg$.
Note that if  $A_{11} \gtrsim \mbox{a few}$, the condition is not satisfied and 
VHE emission will be suppressed for a typical GRB.

\bigskip

\subsection{Radiative efficiency}

The fast-cooling condition for electrons upscattering the X-ray afterglow radiation
can be written in terms of the compactness parameter as
\begin{align}
\frac{\tdyn}{\tIC}
=\frac{4\sigmaT \uX \gamma}{3\me c } \frac{R}{c\Gamma}
= \frac{4}{3} \, \lX  \gamma > 1,
\label{eq:tcool:1}
\end{align}
where 
$\tIC=3m_ec/4\sigmaT\uX\gamma$ is the IC cooling timescale for an electron with
a thermal Lorentz factor $\gamma\sim\gth$ injected at the shock, and
$\tdyn = R/(c\Gamma)$ is the dynamical time 
of the blast wave expansion; both timescales are measured 
in the fluid frame.
Here $\lX$ should be taken as the compactness of the afterglow radiation field below 
the Klein-Nishina energy, which for freshly shocked thermal electrons falls into the XRT band,
\begin{align}
E_{\rm KN} = \me c^2 \frac{\Gamma}{\gth} = 0.46 \,  \left(\frac{\mue}{2}\right)^{-1} \left(\frac{\epse}{0.3}\right)^{-1} \, \mbox{keV},
\label{eq:EKN}
\end{align}
where we have used $\gth = \Gamma\mue\epse\mprot/\me$.

The electrons radiating at energy $x=E/(\me c^2)$ in the lab frame must have
$\gamma \ge x^{\prime} \approx x/\Gamma$, thus the ratio of 
the dynamical and IC cooling timescales satisfies
\begin{align}
\frac{\tdyn}{\tIC} > \frac{4}{3} \, \lX  \frac{x}{\Gamma}.
\label{eq:tcool:2}
\end{align}
Note the similarity between this condition
and Equation~(\ref{eq:opac:gen}) 
for $\gamma\gamma$ opacity.
The numerical coefficients on the right hand sides of the two expressions differ by a factor $\sim 10$, so one can write%
\footnote{The compactness $\lX$ relevant to $\gamma\gamma$ opacity and IC cooling  
        samples slightly different parts of the X-ray spectrum (compare Equations 
        (\ref{eq:Esobs}) and (\ref{eq:EKN})). For $\beta\approx 1$ the difference in 
        $\lX$ amounts to at most a factor of a few and
        and does not invalidate the conclusion of the present discussion.
}
$\tdyn/\tIC > \taugg(x)$,
which brings about an important point: {\it whenever the source is becoming transparent 
to high-energy photons of a certain frequency (i.e. $\taugg(x)\approx 1$),
the electrons emitting those photons are in the fast-cooling regime (i.e. radiatively efficient) 
at that time.} In the case of a wind medium, the VHE luminosity peaks at $\tgg$, and hence
the fast-cooling condition is automatically satisfied.

By contrast, in the ISM case the source is transparent to $\sim 100$~GeV emission
throughout most of its duration,
and condition (\ref{eq:tcool:1}) does yield  
an interesting constraint.
For the electrons injected at $\tdecI$ one obtains
\begin{align}
\frac{\tdyn}{\tIC}
= \frac{4}{3} \, \lX  \gth
\stackrel{\tdec}{=} 5.5 \,  \, \frac{\varepsilon_{{\rm X},-1} \, n^{2/3} \, \E_{{\rm GRB}, 53} \, \Gamma_2^{4/3}}{\E_{{\rm kin}, 53}^{2/3} \Delta_{{\rm X}, 1}} \left(\frac{\epse}{0.3}\right).
\end{align}
This shows that the thermal electrons are radiatively efficient (i.e. fast cooling)
at the deceleration radius
unless the GRB has a particularly weak X-ray afterglow, i.e. $\epsX \ll 0.1$.

\subsection{Inverse Compton spectrum}

The fraction of the energy of the cooling electrons radiated in the observable band 
(between $\sim 50$ and few hundred GeV, see below) depends on the IC spectrum.
The IC spectral slope from fast-cooling thermal electrons roughly mimics the slope of 
the target X-ray radiation if the spectral index of the latter satisfies $\beta > 0$.
The average index of observed X-ray afterglows, $\beta\approx 1$, results in an IC 
spectrum that is approximately flat in $\nu F_{\nu}$, placing a substantial fraction of the 
total energy near $100$~GeV even if the high-energy turnover $\Emax$ is above several TeV.

If the afterglow spectrum is hard ($\beta < 1$), most of the IC power is emitted near the 
cutoff energy $\Emax\approx\Gamma\gth\me c^2$.
This is most relevant if the burst environment is a low-density ISM, in which case 
$\Emax$ can exceed $\sim 100$~TeV in fast jets (Equation \ref{eq:EIC}).
Only a minor fraction of the IC luminosity then emerges near $100$~GeV; 
this fraction can be somewhat increased by reprocessing the highest-energy photons 
to lower energies via pair cascades.

Approximately half of the observed X-ray afterglows have $\beta>1$. 
In these bursts the electrons predominantly cool on the lowest-energy photons available;
then the peak of the IC spectrum depends of the location of 
the low-energy break of the optical to X-ray afterglow spectrum, and most of the IC power could be emitted below the TeV band.

\subsection{Synchrotron losses}

\label{sec:syn}

Another factor that can suppress high-energy emission is synchrotron cooling of the 
TeV-emitting electrons, which is determined by the compactness
\begin{align}
l_B &= \frac{\sigmaT}{\me c^2} \frac{U_B R}{\Gamma} = \nonumber \\
&= 
\left\{ \begin{array}{ll}
	8.0\times 10^{-4} \,  \frac{\displaystyle \varepsilon_{B,-3} \, A_{11}^{1/2} \, \E_{{\rm kin}, 53}^{1/2}}{\displaystyle \E_{{\rm GRB}, 53}^{3/4}},	& \, \mbox{(wind)}	\\
	 4.5\times 10^{-5} \,\, {\displaystyle \varepsilon_{B,-3} \, n^{2/3} \, \E_{{\rm kin}, 53}^{1/3} \Gamma_2^{1/3} }  					& \, \mbox{(ISM),}
	\end{array}
	\right.
\label{eq:lB}
\end{align}
where $\varepsilon_B$ is the fraction of the shock-dissipated energy stored in the 
magnetic field, and the wind and ISM cases have been calculated at $\Rpm$ and 
$\RdecI$, respectively.

The ratio of IC and synchrotron cooling rates is given by
\begin{align}
\frac{\tIC}{\tsyn} = \frac{l_B}{\lX} = 
\zeta\, \Delta_{\rm X, 1} \, \frac{\varepsilon_{B,-3} }{\varepsilon_{{\rm X},-1}} \, \frac{\E_{{\rm kin}, 53}}{\E_{{\rm GRB}, 53}} ,
\label{eq:tsyn}
\end{align}
where the numerical factor $\zeta = 0.1$ (wind) or $0.6$ (ISM). 

If $\tsyn<\tIC$, the electrons cool predominantly by synchrotron rather the IC emission;
the IC luminosity is suppressed by the factor $\tsyn/(\tIC + \tsyn)$.
If the X-ray afterglow light curve follows $\LX\propto t^{-1}$,
the ratio of synchrotron and IC cooling rates is independent of time.
Thus one typically requires $\epsB \lesssim 10^{-3}$ for the
synchrotron losses to be insignificant.
This is a reasonable expectation given that efficient particle acceleration required for 
producing the X-ray (synchrotron) afterglow is suppressed if the upstream magnetization 
$\epsB\gtrsim 10^{-3}$ \citep{SironiSpitkovsky2011}.
While the self-generated small-scale stochastic magnetic
field near the forward shock could be significantly stronger,
it tends to decay on a scale comparable to the plasma skin depth
(e.g. \citealt{Chang2008,Lemoine13}; however, see also \citealt{Keshet2009}),
which is much shorter than the IC or synchrotron cooling length in the downstream.


\section{Detecting TeV emission}

\label{sec:det}

\subsection{Response time}

Due to their narrow field of view, Imaging Atmospheric Cherenkov Telescopes (IACT) rely on an independent trigger from a wide-field instrument
such as {\it Swift}/BAT or {\it Fermi}/GBM for 
GRB alerts. IACT observation of the GRB will be delayed by a few factors:
obtaining the burst position via GCN, observer response, and 
the telescope slew time.

As described in Section~\ref{sec:TEVemiss},
the onset of VHE emission must be delayed with respect to the prompt trigger by
pair-loading, $\gamma\gamma$ opacity, and the time 
it takes the blast wave to receive its energy from the jet.
The bulk of VHE radiation emerges over minutes to hours and can last for up to a day.
The required response times are
easily accessible to current as well as next generation IACT.
For example, the VERITAS instrument responded to 36 burst alerts between 2006 and 2013,
and for half of them the data taking started before 180 s after the trigger \citep{Inoue2013}.
For 16 burst triggers followed up by
MAGIC (Major Atmospheric Imaging Cerenkov Telescope)
between 2005 and 2008 the observation started within 180 s in half of the cases
\citep{Garczarczyk2008}.

The Large Size Telescopes that  provide the
low-energy sensitivity ($20-200$~GeV) of the Cherenkov Telescope Array (CTA)
will have a projected azimuthal slew time of $180$ degrees in $20$ seconds
\citep{Inoue2013}, comparable to that of MAGIC.

\subsection{Sensitivity}

With $0.5$~hour integration and favorable observing conditions,
current IACT can achieve a fluence limit at $100$~GeV that is
several orders of magnitude deeper than LAT,
and over an order of magnitude deeper than {\it Fermi}/LAT throughout its entire bandpass.
On average, LAT detects $\sim 10$\% of GRBs in its field of view, with a
typical fluence of $\sim 10-20$\% of the prompt MeV radiation \citep{Ackermann2013}.
The higher sensitivity of IACT suggests that they could detect 
a similar fraction of observed bursts if the output above $\sim 100$~GeV constitutes 
only $\sim 1\%$ of the prompt GRB energy.
This level of VHE emission is in the theoretically expected range. 
The blast wave radiates at least a few tens of percent of its energy (\Eq~\ref{eq:tLIC}),
and the analysis in Section~\ref{sec:TEVemiss} shows that a significant fraction $\epsTeV$ 
of this radiation is emitted in the VHE band.

As an example, consider a GRB at redshift $z=1$, with the isotropic equivalent of the
blast-wave energy $\Ekin = 10^{53}$~erg~s$^{-1}$.
The observed fluence of radiation emitted by the blast wave (\Eq~\ref{eq:tLIC}) is given by
\begin{align}
{\cal F}
\sim 3.4\times 10^{-6} \,
\E_{{\rm kin},53}  \left(\frac{\epse}{0.3} \right)
\left(\frac{\chi}{0.3} \right)
\left( \frac{\dL}{6.7 \,\mbox{Gpc}} \right)^{-2}
\, \mathrm{erg} \, \mathrm{cm}^{-2},
\label{eq:flux}
\end{align}
where the luminosity distance $\dL$ is normalized to its value at $z=1$ in standard cosmology.
The VHE photon count at the detector is then given by
\begin{align}
{\cal N} &\sim  \A\, \frac{\epsTeV{\cal F}}{\overline{E}} \approx
700  \,
 \E_{{\rm kin},53}  
 \left(\frac{\epse}{0.3} \right)
 \left(\frac{\chi}{0.3} \right) \,
\left( \frac{\epsTeV}{0.1} \right) 	\nonumber \\
&\times \left(\frac{\A}{5\times 10^8 \, \mathrm{cm}^2}\right)
\left( \frac{\overline{E}}{150\, \mathrm{GeV}} \right)^{-1}
\left( \frac{\dL}{6.7 \,\mbox{Gpc}} \right)^{-2}
\label{eq:count}
\end{align}
where $\A$ is normalized to the effective area of the VERITAS instrument at $150$~GeV,
and $\epsTeV$ is the energy fraction that is carried by photons arriving with energies around
$150$~GeV and avoiding absorption by the extragalactic background light (EBL).

\begin{figure}[t]
  \begin{center}
  \includegraphics[width=0.47\textwidth, trim=0 0 0 1cm]{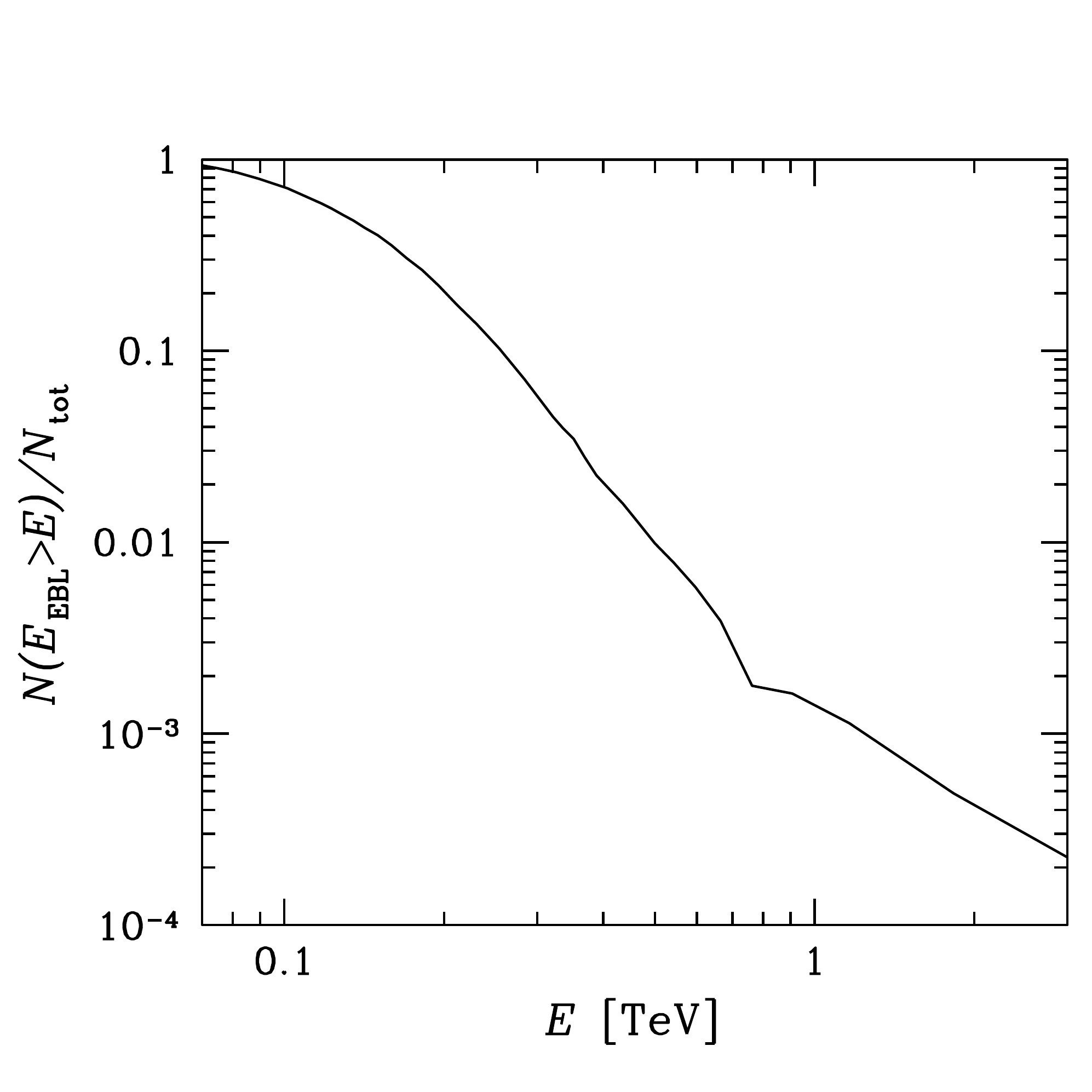}
  \vspace*{-0.4cm}
  \end{center}
  \caption{Fraction of bursts 
  observable at photon energy $E$ without strong attenuation by the EBL.
  Only bursts detectable by \Swift are considered.
  The redshift and prompt energy distributions are taken form \citet{Kanaan2013},
  and the model for EBL attenuation is from \citet{Dominguez2011}.
  }
  \label{Fig:E_EBL}
\end{figure}

The chosen ``average'' photon energy, $\overline{E} = 150$~GeV,
corresponds to the relatively narrow 
spectral window which maximizes the probability for detection. This window is  shaped by
(1) the effective area of Cherenkov detectors which drops steeply below $\sim 100$~GeV,
and
(2) the $\gamma\gamma$ attenuation by the EBL, which increases rapidly above 
$E_{\rm EBL}\sim 100-200$~GeV.
Here the characteristic photon energy $E_{\rm EBL}$ is defined so that the optical 
depth to absorption $\tau_{\rm EBL}(E_{\rm EBL})=1$. The EBL role is illustrated by
Figure~\ref{Fig:E_EBL}, which shows the fraction of {\it Swift}-detectable bursts with
$E_{\rm EBL}$ above a given photon energy $E$. Only a small fraction of bursts with  
low redshifts could in principle be observed above a few hundred GeV.

Given the high effective area of the detector and the moderate attenuation by 
the EBL at $E\sim 150$~GeV, one could expect a significant detection rate of GRBs 
by VERITAS.
Based on the approximate VERITAS background count rate $\sim 1-10$~min$^{-1}$,
a $5\sigma$ detection over $1$~hr would require $\sim 50-100$ source counts.
According to Equation~(\ref{eq:count}), even bursts with a comparatively modest energy 
budget can satisfy this requirement. The absence of any detections so far indicates 
significant suppression of the VHE emission below the estimate given in \Eq~(\ref{eq:count}).
Strong suppression of IC emission by synchrotron losses appears unlikely for 
plausible $\epsB$. The most obvious candidate for this suppression is a
high intrinsic opacity to $\epm$ pair production, which would suggest that GRBs explode 
into dense environments (Equation \ref{eq:yes}).

It is evident from Figure \ref{Fig:E_EBL} that owing to EBL absorption,
the prospects of GRB detection by ground-based gamma-ray instruments
rely strongly on their sensitivity at $\lesssim 0.1$~TeV.
The ability to detect lower energy photons also helps to avoid {\it intrinsic} absorption,
making it easier to satisfy condition (\ref{eq:yes}). In this respect
the next generation Cherenkov Telescope Array (CTA) will be a significant improvement
over currently operating observatories.
For example, the projected effective area of CTA South is $10^8$~cm at $30$~GeV,
$\sim 3$ orders of magnitude higher than 
that of VERITAS (for which $30$~GeV is very near the energy threshold), and
approximately four orders of magnitude higher than {\it Fermi} LAT in the same range.

Note also that the most energetic GRB photon detected by LAT so far 
(in GRB~130427A) had the energy of 95~keV and arrived $243$~s after the trigger
\citep{Ackermann2013}.
This corresponds to thousands of expected counts in current Cherenkov telescopes.
While VERITAS did indeed observe GRB 130427A, the observation started only 
$\sim 20$~hr after the trigger and resulted in a non-detection \citep{Aliu2014}.
most likely because the VHE emission had already ceased ($t>t_{\rm 100 GeV}$).


\section{Monte-Carlo modeling of detection rates}

\label{sec:MC}

To make quantitative predictions for the detection rates of GRBs with Cherenkov instruments,
we have simulated a large number of bursts spanning a range of redshifts and intrinsic 
energies. For each simulated burst we calculated the theoretical TeV light curve
as described in Section~\ref{sec:TEVemiss}, taking into account
the evolution of the high-energy spectrum, intrinsic absorption etc. 
Then the flux received by the detector is corrected for the cosmological redshift of 
the burst and EBL absorption.
Finally, the number of detector counts and detection significance is found based on the effective area and background count rate of the relevant instrument.

The following information is required as an input for the Monte Carlo simulation:
\begin{enumerate}
\item
Isotropic energy equivalent and redshift distributions of GRBs. We use the results from
\citet{Kanaan2013}, derived from the {\it Swift}/BAT catalog.
\item
Distributions of X-ray afterglow energetics and spectral slopes.
This information is required for determining the opacity of the source to 
VHE radiation, as well as the radiative efficiency of the blast wave in the TeV band.
\citet{Margutti13} find a strong correlation between the prompt and X-ray afterglow 
energy $\EX$; we use this to draw $\epsX$ such that $\EX = \epsX \EGRB$, 
approximate the X-ray light curve as $\LX = \EX/(t \DX)$,
and use $\DX = 10$ as the average duration of the afterglow in logarithmic units.
The distribution of X-ray spectral slopes is taken from \citet{Grupe2013};
for long GRBs it has a mean $\beta \approx 1$ and standard deviation 
$\sigma_{\beta} \approx 0.43$.
\item
Model for EBL absorption. We use the gamma-ray opacities given by
\citet{Dominguez2011},
as functions of redshift and photon energy.
In the relevant energy range ($\sim 30-1000$~GeV)
the attenuation factors $e^{-\tau}$ from \citet{Dominguez2011}
agree with those found in
\citet{Franceschini2008} within a few percent,
and are within a factor $\sim 2$ of the results of
\citet{Gilmore2012}.
\item
Effective area and background rate of the detector,
at $20^{\degree}$ zenith angle.
The specifications for Cherenkov Telescope Array and VERITAS
are taken from the CTA consortium website\footnote{https://www.cta-observatory.org}
and VERITAS website\footnote{http://veritas.sao.arizona.edu}, respectively.
\end{enumerate}
In addition to the above, one must specify 
the external medium density $A$, the jet Lorentz factor $\Gjet$,
and the radiative efficiency of the burst, which 
determines the explosion energy (isotropic equivalent) after the prompt GRB, 
$\Ekin = \EGRB (1-\epsrad)/\epsrad$.
Fortunately, the results turn out to be insensitive to the choice of the poorly known $\Gjet$;
we choose $\Gjet =100$ in most of our simulations.
The main adjustable parameters of the model are $A$ and $\epsrad$.

The Monte Carlo procedure for each burst is as follows:
\begin{enumerate}
\item
Draw the redshift and $\EGRB$ from their distributions.
Calculate $\tpm$,
the time at which $e^\pm$ loading ceases.
\item
Draw the energy and spectral index of the X-ray afterglow. 
\item
Integrate the blast wave evolution. We use simplified
energy and momentum conservation equations for the blast wave dynamics which
treat the shocked jet material as a single body.
The calculation includes the early stage of the evolution when
energy is still being supplied to the blast wave via the reverse shock.
Radiative losses are taken into account; as long as the electrons are fast-cooling, 
a fraction $\epse$ of the shock-dissipated energy is lost to radiation.
\item
At each time step (at $t>\tpm$) perform the following calculations.
(1) Determine the Lorentz factor of the blast wave
$\Gamma$ and electron injection energy $\gth$.
(2) Determine the emitted high-energy spectrum using a library of IC spectra.
The library is calculated beforehand using Monte-Carlo radiative transfer simulations 
for fast-cooling electrons in a relativistic blast wave.
The main parameter determining the high-energy spectral shape is
the spectral index of the target X-ray radiation;
neither $\Gamma$ nor $\gth$ influence the shape of the IC spectrum.
(3) Based on the afterglow spectrum, calculate the opacity of the source to gamma-rays 
as a function of frequency, and determine whether the electrons are in the fast-cooling 
regime. Attenuate the IC spectrum by $e^{-\taugg(x)}/(1+\tIC/\tdyn)$.
The result of this stage is
the high-energy spectrum released to the observer from the current radius.
\item
Calculate the fluence spectrum at the observer accounting for EBL absorption, the 
predicted counts in the detector, and the detection significance 
using the method described in \citet{Li1983}.
\end{enumerate}


\section{Results and discussion}

\label{sec:res}

\subsection{Wind medium}

\begin{figure} 
  \begin{center}
  \includegraphics[width=0.47\textwidth, trim=0 0 0 1cm]{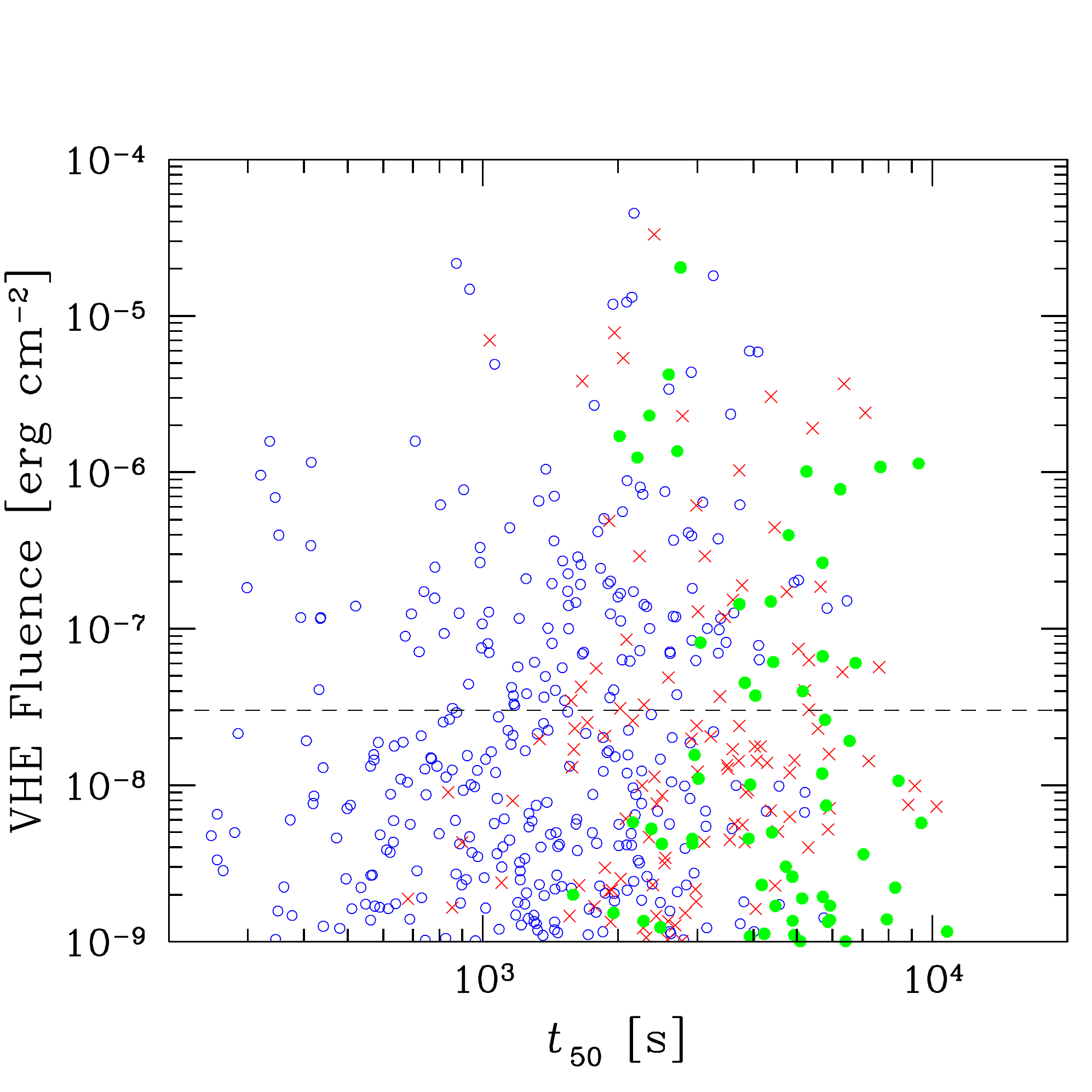}
  \vspace*{-0.4cm}
  \end{center}
  \caption{The distribution of simulated bursts (occurring over one year)
  in the plane of two main observational parameters: 
  fluence above $50$~GeV and accumulation time $t_{50}$ of $50$\% of the VHE fluence.
  Different symbols correspond to three different density parameters of 
  the external wind medium:
  $A=10^{11}$g~cm$^{-1}$ (blue open circles), 
  $3\times 10^{11}$g~cm$^{-1}$ (red crosses), and $5\times 10^{11}$g~cm$^{-1}$ (green filled circles).
  The radiative efficiency of the prompt emission is fixed at $\epsrad=0.5$ in 
  all three simulations. The choice of the jet Lorentz factor $\Gjet\simgt 100$ weakly 
  affects the results and is fixed at $\Gjet=100$.
  The horizontal dashed line
   ${\cal F}=3\times 10^{-8}$~erg~cm$^{-2}$
  indicates the CTA sensitivity; it
  approximately corresponds to the fluence limit of CTA South
  between 50 and 200~GeV for 0.5~h integration.
  The number of simulated bursts with $\FlTeV$ above this value is 128, 36 and 21 in   
  the increasing order of $A$.
  }
  \label{Fig:Fl_tobs}
\end{figure}

\begin{figure} 
  \begin{center}
  \includegraphics[width=0.47\textwidth, trim=0 0 0 1cm]{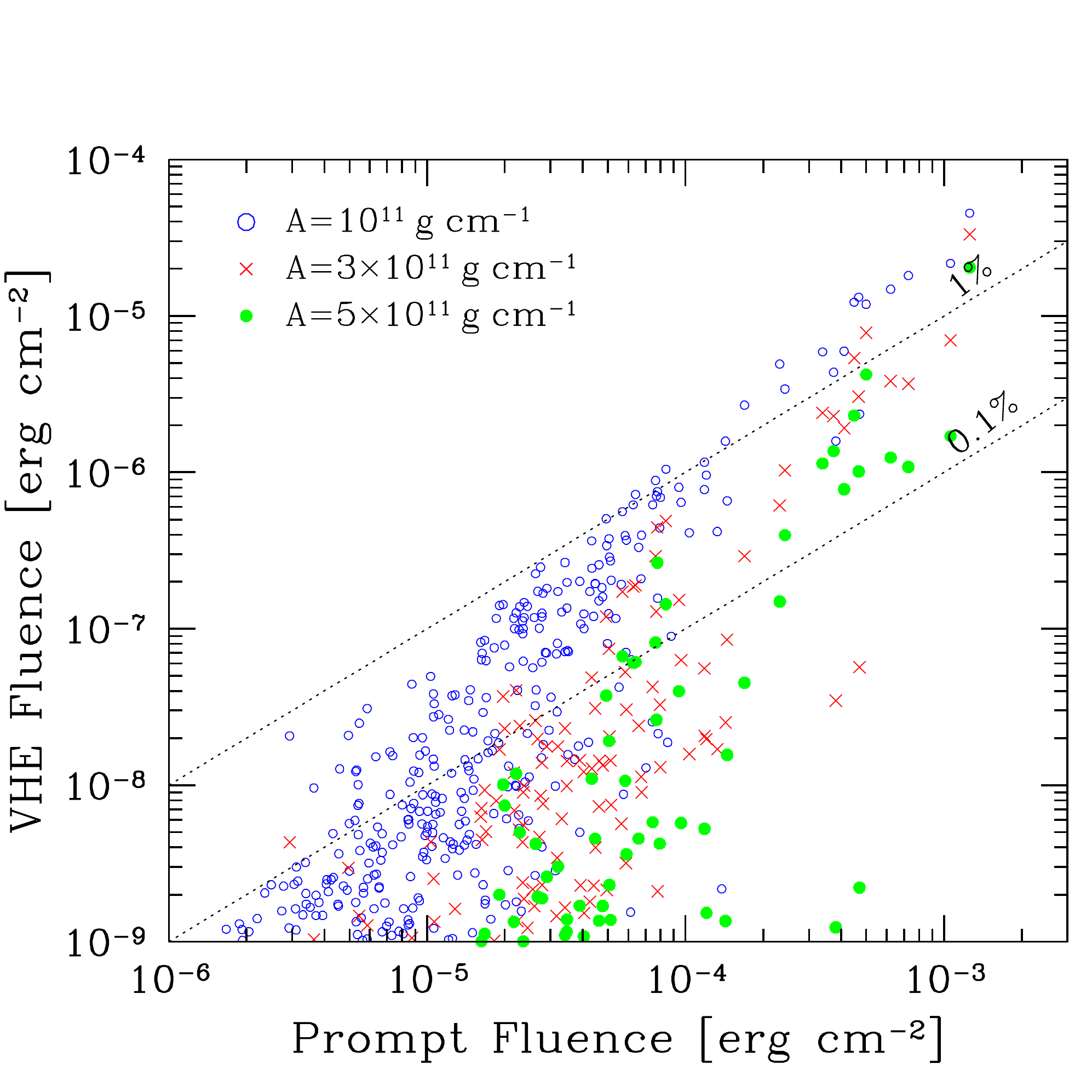}
  \vspace*{-0.4cm}
  \end{center}
  \caption{The same set of simulated bursts as in Figure \ref{Fig:Fl_tobs},
  viewed in the plane of prompt MeV fluence vs. VHE fluence.
  }
  \label{Fig:Fl_Fl}
\end{figure}

The main parameters that determine detectability of a GRB by IACT are the TeV 
fluence and the typical arrival time of the VHE photons.
Figures \ref{Fig:Fl_tobs} and \ref{Fig:Fl_Fl} show a 
Monte Carlo realization of GRBs exploding into a wind medium, accumulated over  
one year of observations.
The typical photon arrival times range from a few minutes to a few hours,
and are significantly delayed with respect to $\tpm\sim 100$~s.
The delay is caused by the source being
initially opaque to photons in the TeV band in most cases;
the high-energy flux peaks right after the optically thin transition at $\tgg$,
and continues until $\sim t_{\rm 100 GeV}$ (Eqs. \ref{eq:t100} and \ref{eq:tgg}).

The delayed $\gamma\gamma$ transparency is the reason why the TeV emission typically peaks later in denser winds. In models with higher density parameter $A$ 
the optically thin transition is delayed due to a lower Lorentz factor of the blast
wave (it decelerates earlier). On the other hand, the production of TeV gamma-rays 
ceases earlier (see Equations \ref{eq:t100}, \ref{eq:tgg} and \ref{eq:yes}). As a result,
the number of bursts above a given fluence threshold decreases with increasing 
$A$ (Figure \ref{Fig:Fl_tobs}).

The VHE fluence is compared with the prompt GRB fluence for the simulated bursts
in Figure \ref{Fig:Fl_Fl}.
There is an expected positive correlation between the prompt and VHE fluences.
The $\FlTeV$-$\FlGRB$ relation is steeper than linear,
which reflects the fact that less energetic blast waves are more opaque to 
$\gamma\gamma$-absorption in the VHE producing stage.
The predicted VHE fluence is between $\sim 0.1$-$1$~\% of the prompt fluence
for most bursts with detectable TeV emission.
 
\begin{figure} 
  \begin{center}
  \includegraphics[width=0.47\textwidth, trim=0 0 0 1cm]{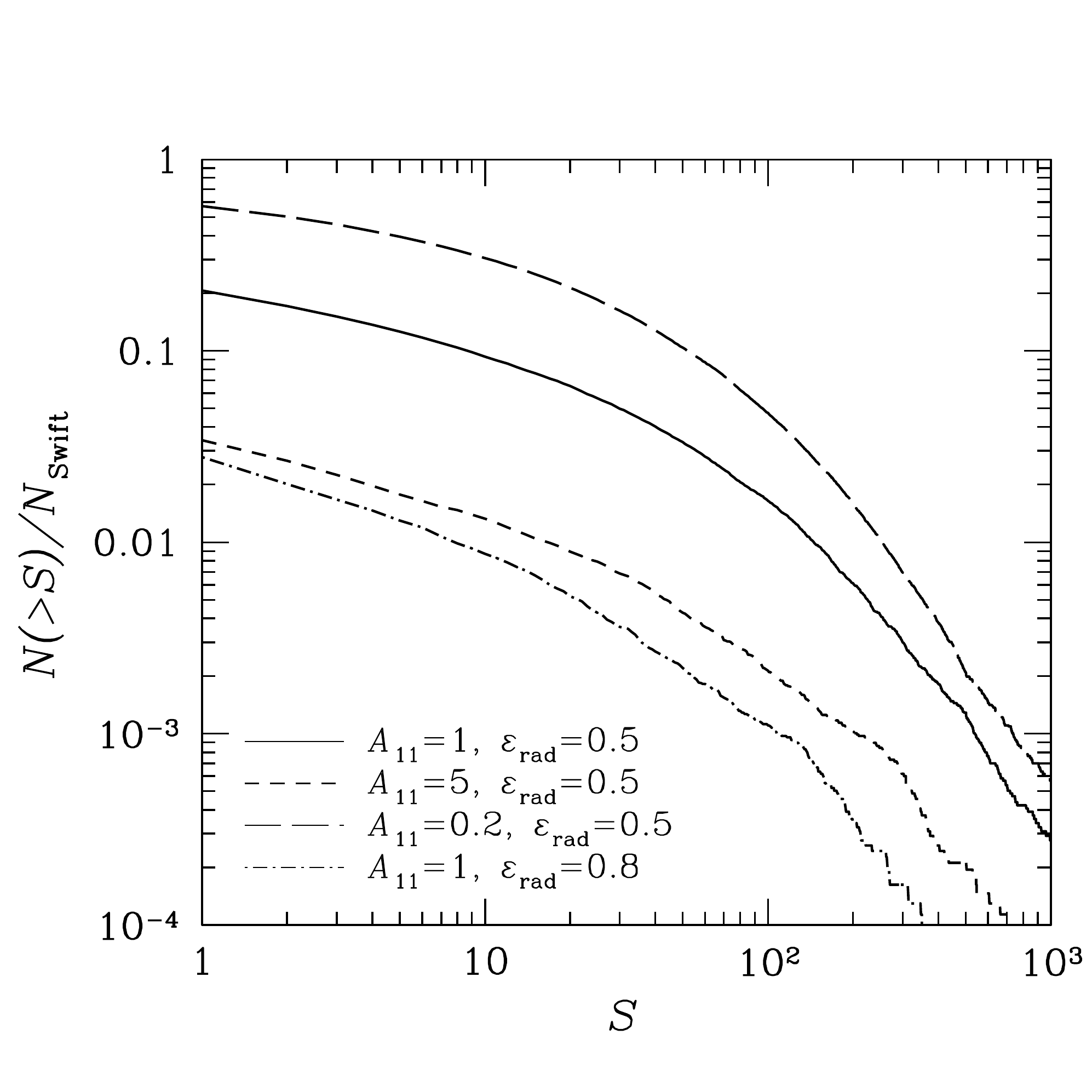}
  \vspace*{-0.4cm}
  \end{center}
  \caption{Fraction of {\it Swift}-detectable bursts that are
  also detectable by CTA, according to the detection significance $S$.
  A response time of $100$~s with respect to the prompt trigger is assumed.
  The rates are not corrected for suppression due to
  viewing angle/earth occultation effects, instrument duty cycle, weather conditions etc.
  }
  \label{Fig:CTA:signif}
\end{figure}

\begin{figure} 
  \begin{center}
    \includegraphics[width=0.47\textwidth, trim=0 0 0 1cm]{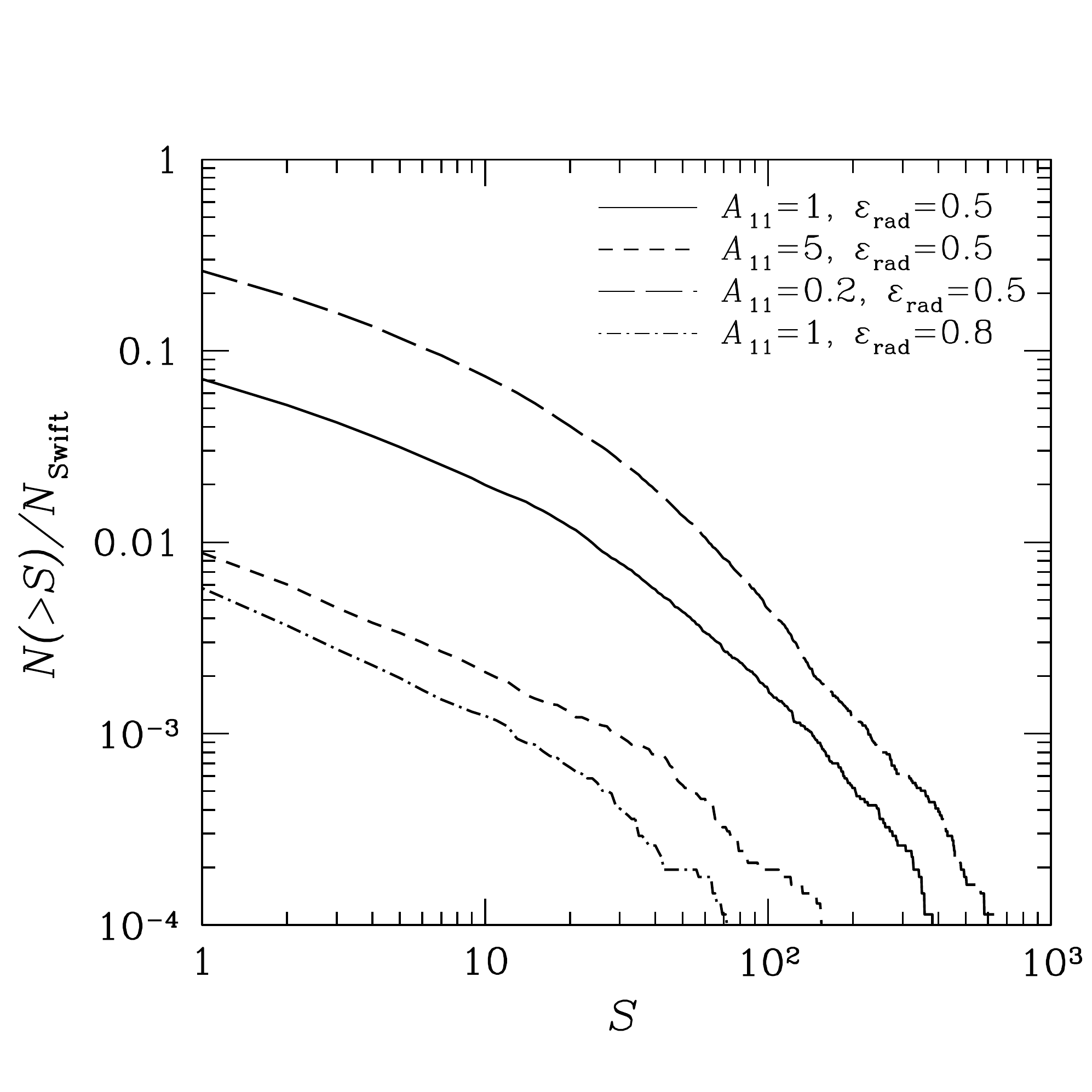}
  \vspace*{-0.4cm}
  \end{center}
  \caption{Same as Figure \ref{Fig:CTA:signif}, but for VERITAS.
  }
  \label{Fig:VER:signif}
\end{figure}

The fraction of {\it Swift}-triggered bursts that are detectable by CTA or VERITAS is shown 
in Figures \ref{Fig:CTA:signif} and \ref{Fig:VER:signif}, as a function of detection significance.
Depending on $\epsrad$ and $A$, up to a few tens of percent of \Swift bursts
are detectable with high significance.
The detectable fraction is smaller for bursts with lower jet kinetic energies 
(i.e. higher $\epsrad$) exploding into a dense medium.
The number of potentially detectable bursts is $\sim 3$ times higher
for CTA compared to VERITAS.
The fact that VERITAS has not detected any of the 36 bursts observed by
2013 is consistent with dense winds $A \gtrsim 10^{11}$~g~cm$^{-1}$
and/or high radiative efficiencies at the prompt phase $\epsrad > 0.5$.

In Figures \ref{Fig:Fl_tobs}-\ref{Fig:VER:signif} we have used a constant $\Gjet=100$ 
for all simulated bursts, a simplification that mostly reflects our
limited knowledge of GRB jet Lorentz factors.
However, in dense wind environments the blast wave is already in the self-similar 
deceleration regime
by the time most of the TeV radiation is emitted, having ``forgotten'' its initial $\Gjet$.
On the other hand, $\Gjet$ does influence the radiative losses of the ejecta in the early 
pair-dominated stage of the afterglow, and thus affects the blast wave energy available 
at the later, TeV-producing stage.
This effect will be discussed in more detail in Section \ref{sec:radloss} below.

Let us return briefly to the apparent tension between
Equation (\ref{eq:count}) and the non-detections in the VHE band.
A burst with $\EGRB=10^{53}$~erg~s$^{-1}$ at $z=1$ generates a fluence
${\cal F} \approx 3.7\times 10^{-5}$~erg~cm$^{-2}$.
In the count estimate (\ref{eq:count}),
approximately 1\% of this fluence was assumed to be radiated in the TeV band.
Comparing this to Figure \ref{Fig:Fl_Fl}, one finds that the TeV typically constitutes
close to 1\% of the prompt fluence only if $A\lesssim 10^{11}$~g~cm$^{-1}$,
assuming a moderate prompt efficiency $\epsrad\approx 0.5$.
In denser winds the blast wave is significantly less luminous in the VHE band,
due to strong suppression by $\gamma\gamma$-absorption.
The TeV fraction is even smaller in less energetic bursts,
which compensates for their greater numbers
(recall that the role of opacity is controlled by $\Ekin/A$, according to Equation (\ref{eq:yes})).
Overall, wind densities of $A\gtrsim 10^{11}$~g~cm$^{-1}$ seem to be required
to reconcile the lack of detections with theory,
in a agreement with the conclusion drawn from Figure \ref{Fig:VER:signif}.

\subsection{ISM}

The wind of a massive progenitor is the most natural
GRB environment, however a low density ambient medium such as the ISM has not 
been conclusively ruled out for all bursts. Using the GeV emission as a diagnostic, 
it was shown in \citet{Hascoet15} for a small sample of seven energetic GRBs with 
good LAT data that the wind environment is required to reproduce both the bright early 
flash as well as the details of the extended GeV light curve. 
The expected difference in VHE emission between the wind and ISM cases
will become particularly important for CTA observations, and they can 
establish a wind medium for a large sample of bursts.

At radius $\Rpm$ where the forward shock starts producing TeV photons the 
ISM density is approximately $2-3$ orders of magnitude lower than that of the 
Wolf-Rayet wind (note that $\Rpm$ is independent of the external medium density).
This results in a few qualitative as well as quantitative differences between the two cases:
\begin{enumerate}
\item
The blast wave in the ISM expands to a larger radius than in the wind
before transferring a significant fraction of its energy to the external medium
at $R \sim \RdecI$.
The blast wave is luminous in the VHE band at the deceleration radius, since 
$\RdecI > \Rpm$ and the shock-heated thermal electrons are radiatively efficient.
The blast-wave Lorentz factor is high at $\RdecI$, $\Gamma \sim \Gjet$, and
a significant fraction of the dissipated power emerges as multi-TeV radiation, extending to 
$\sim 100$~TeV for jets with highest $\Gjet$. In contrast, $\RdecW < \Rpm$ in a wind 
medium, thus $\Gamma < \Gjet$ at the TeV onset, and the spectrum cuts off at lower energies.

\item
Higher characteristic radii and Lorentz factors in the ISM case result in lower opacities 
to $\gamma\gamma$ pair production; the source is transparent to $0.1-1$~TeV 
gamma rays at $\tdecI$ where the TeV luminosity peaks.
In a wind medium, the VHE gamma-rays are initially blocked in most GRBs;
the high-energy emission then peaks at the optically thin transition.
In dense winds, this transition occurs after the TeV production has already ceased,
and no VHE emission escapes.

\item
In a dense medium the jet can lose a substantial fraction of its kinetic energy at an 
early radiative stage when the pair loading is still high (Section \ref{sec:radloss}).
When $\Zpm \gtrsim 3000$, the bulk of the shock-dissipated energy
is transferred to $\epm$ pairs, and is promptly radiated in the MeV-GeV domain.
In the ISM these early losses amount to a minor fraction of the energy budget,
leaving most of the jet kinetic energy available for dissipation at a later stage when 
most of the TeV radiation is produced.
\end{enumerate}

\begin{figure} 
  \begin{center}
  \includegraphics[width=0.47\textwidth, trim=0 0 0 1cm]{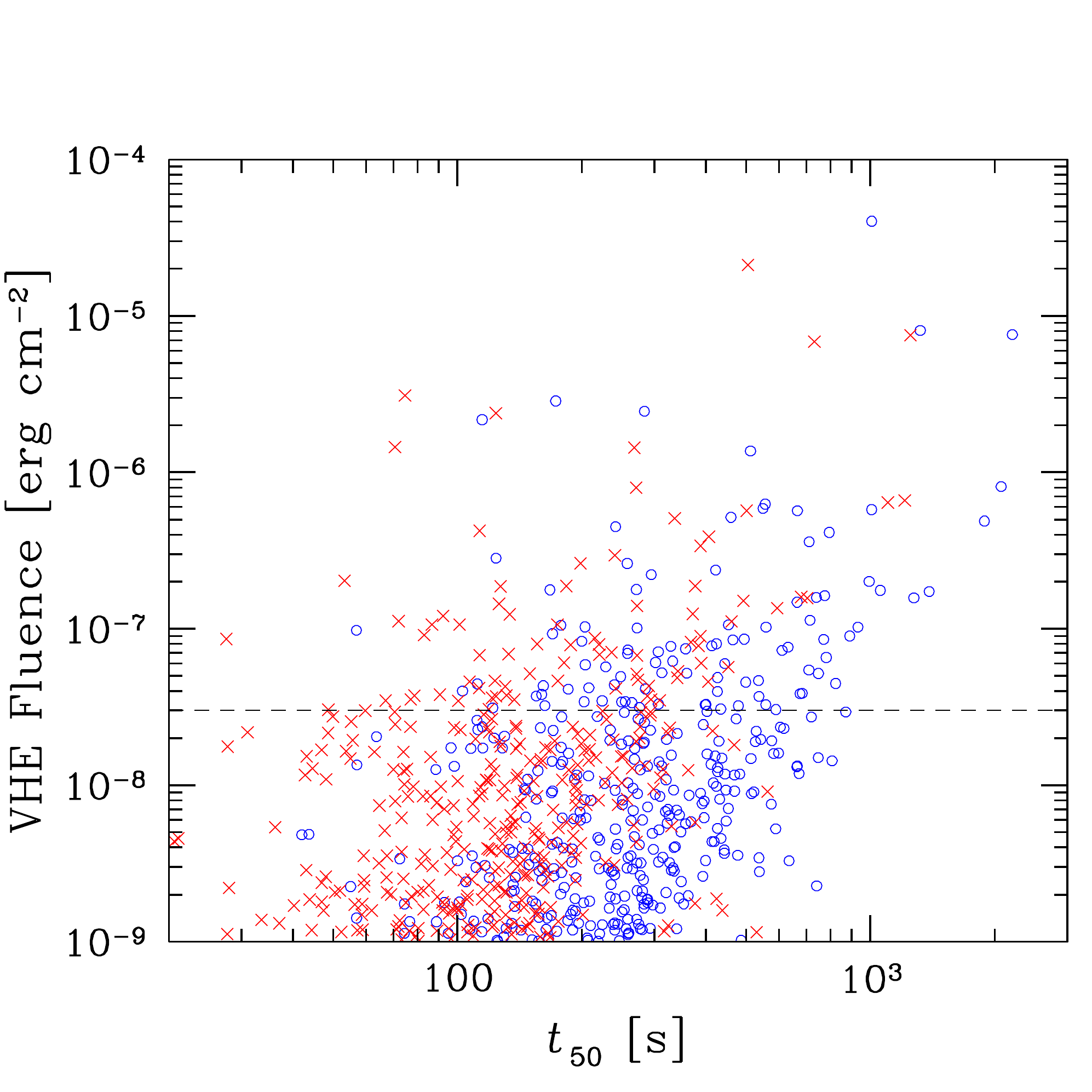}
  \vspace*{-0.4cm}
  \end{center}
  \caption{The distribution of simulated bursts exploding into the ISM, according to their fluence above $50$~GeV
  and accumulation time of $50$\% of the VHE fluence.
  The calculation period is $0.1$ years (in contrast to 1~year in Figures \ref{Fig:Fl_tobs} and \ref{Fig:Fl_Fl}).
  Blue open circles and red crosses correspond to ISM densities $n=1$~cm$^{-3}$ and $n=10$~cm$^{-3}$, respectively.
  Other parameters: radiative efficiency $\epsrad=0.5$, jet Lorentz factor $\Gjet=100$ (fixed).
  The number of simulated bursts with $\FlTeV>3\times 10^{-8}$~erg~cm$^{-2}$ (horizontal dashed line) is 97 and 80 in 
  the increasing order of $n$.
  }
  \label{Fig:ISM:Fl_tobs}
\end{figure}

\begin{figure} 
  \begin{center}
  \includegraphics[width=0.47\textwidth, trim=0 0 0 1cm]{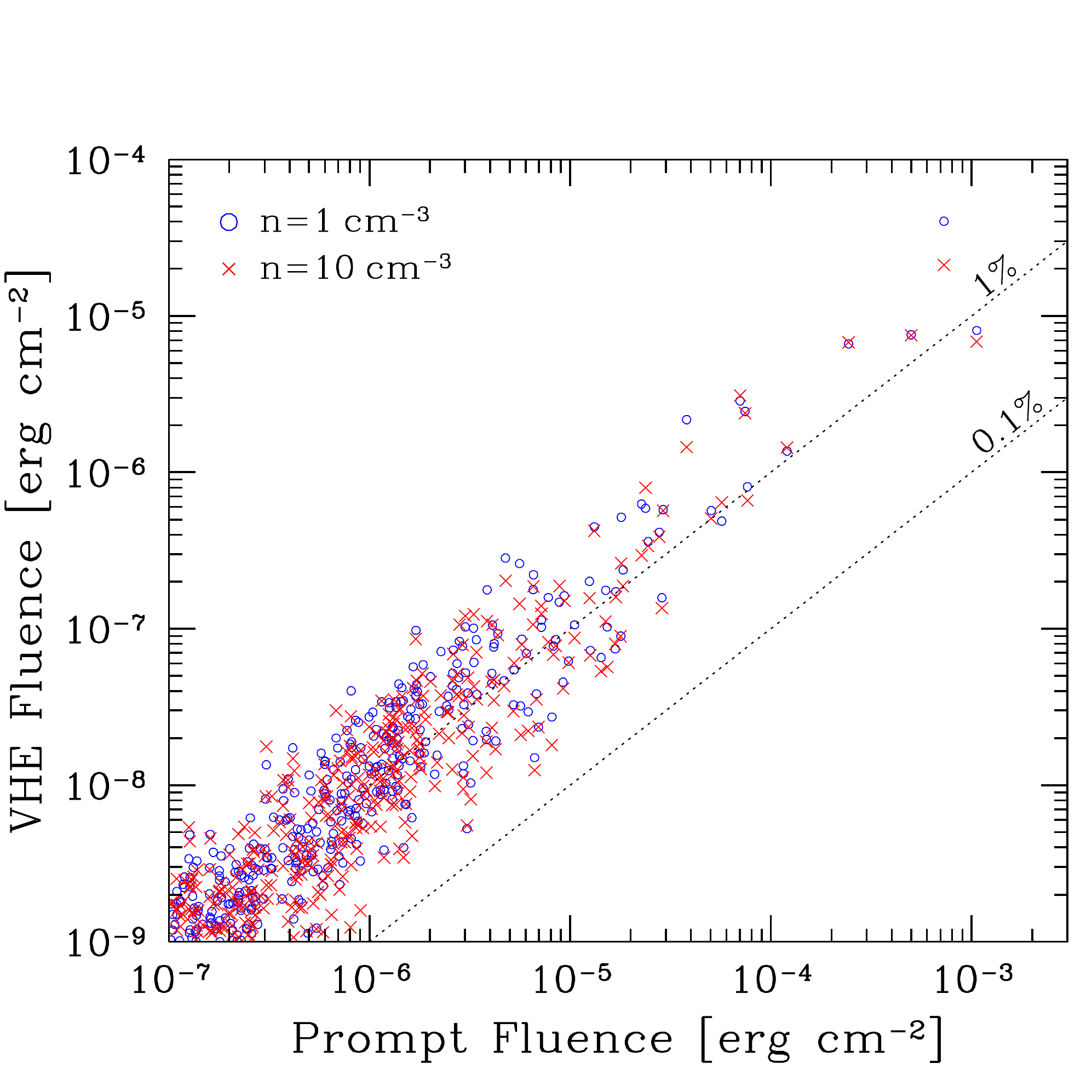}
  \vspace*{-0.4cm}
  \end{center}
  \caption{The same bursts as in Figure \ref{Fig:ISM:Fl_tobs}, according to their prompt and TeV fluences.
  }
  \label{Fig:ISM:Fl_Fl}
\end{figure}

\begin{figure} 
  \begin{center}
  \includegraphics[width=0.47\textwidth, trim=0 0 0 1cm]{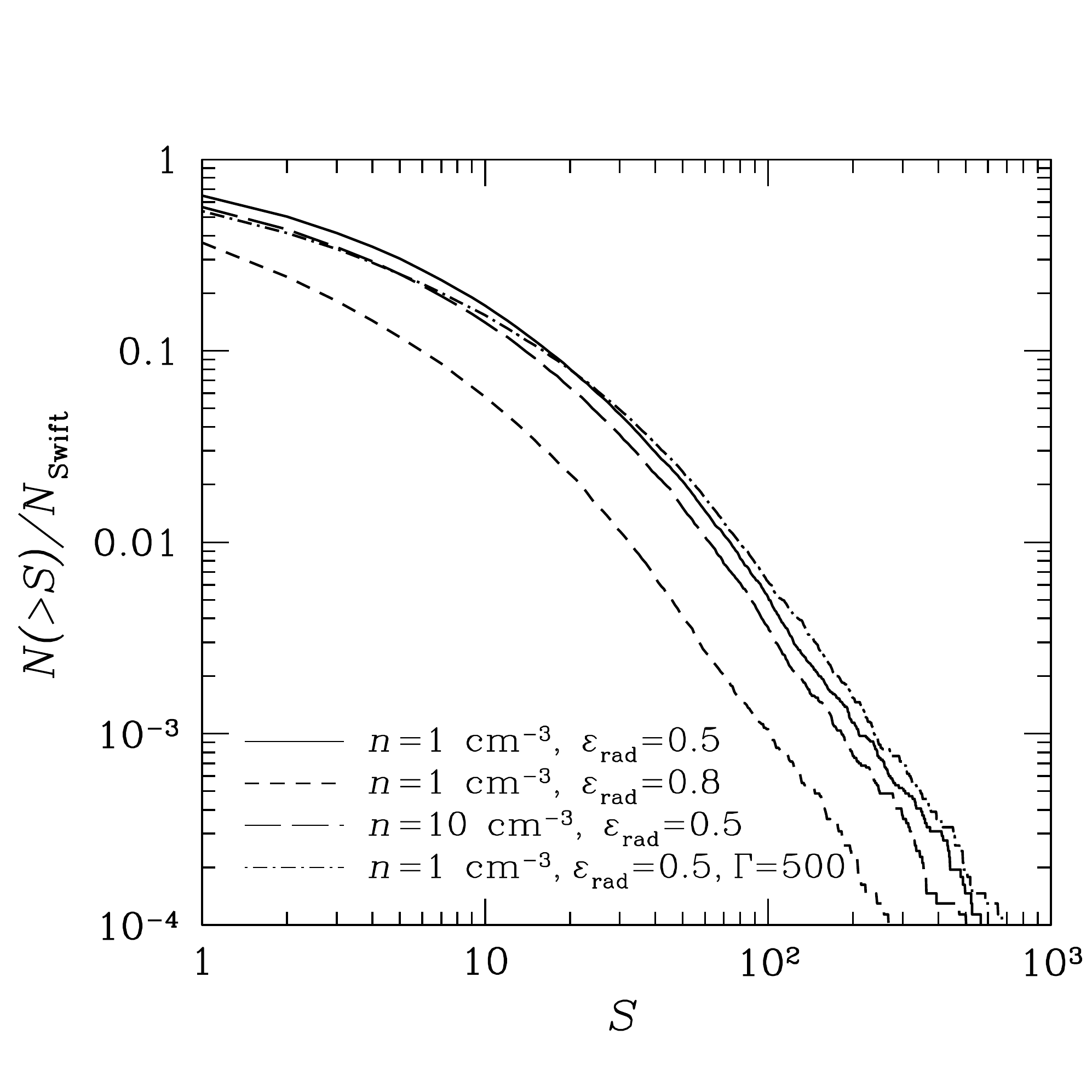}
  \vspace*{-0.4cm}
  \end{center}
  \caption{Fraction of {\it Swift}-detectable bursts 
  that are also detectable by VERITAS, for 
  a constant density interstellar medium.
  A response time of $100$~s with respect to the prompt trigger is assumed.
  }
  \label{Fig:VER:ISM:signif}
\end{figure}

A Monte Carlo realization of bursts exploding into the ISM over 0.1 year is shown 
in Figures \ref{Fig:ISM:Fl_tobs} and \ref{Fig:ISM:Fl_Fl}. The number of
bursts above any given fluence threshold
is significantly higher than in the wind case\footnote{Note that
          the scatter plots for the wind case were calculated for a period of 1 year, 
          in contrast to $0.1$ years in the ISM case.}.
Another notable difference is the systematically earlier peak of the TeV radiation,
which in the ISM is controlled by the dissipation rate at the forward shock
(which peaks at $\Rdec$), rather than $\gamma\gamma$ opacity as in the wind case.
Increasing the ambient density initially shifts the TeV peak to earlier times 
(Figure \ref{Fig:ISM:Fl_tobs}), owing to earlier deceleration.
At even higher densities the opacity effects eventually become important,
delaying the TeV peak similarly to the wind case.

The TeV fluence constitutes $\sim 1$\% of the prompt fluence for the parameters in Figure \ref{Fig:ISM:Fl_Fl}.
The approximately linear $\FlTeV$-$\FlGRB$ relation shows that intrinsic 
$\gamma\gamma$ opacity does not play a significant role,
which would lead to a steeper dependence similar to the wind case.

Figure~\ref{Fig:VER:ISM:signif} shows the fraction of bursts in the ISM that trigger \Swift
and are also detectable by VERITAS. Compared to the wind case,
the number of TeV-detectable bursts is substantially higher.
This result is unsurprising in light of the preceding discussion, if one recalls that
(1) a few tens of per
cent of the blast wave energy is transferred to TeV-emitting electrons ($\epse \approx 0.3$),
(2) the electrons are fast-cooling,
and most importantly (3) the forward shock is transparent to TeV gamma-rays when the forward shock luminosity peaks at $\Rdec$.

In three of the four models shown Figure~\ref{Fig:VER:ISM:signif} we chose $\Gjet=100$.
The choice of $\Gjet$ was not important in the wind case (because the VHE emission 
peaked at $R>\Rdec$), however it is more important in the ISM case, since $\Gjet$ 
determines $\tdec$ at which the emission peaks. An early peak at high $\Gjet$
could easily be missed by TeV instruments.
On the other hand, a 
high $\Gjet$ increases the fraction of the IC energy 
emitted in the TeV band ($E_{\rm IC,max}\propto\Gamma^2$, Equation (\ref{eq:EIC})).
Note also that the duration of VHE emission does not depend on $\Gjet$ (Equation \ref{eq:tend:ISM}),
so the burst could still be detected even if the peak was missed.
We have tested how increasing $\Gjet$ 
to 500 would affect our results and found that the 
change in the detection rate is small (Figure \ref{Fig:VER:ISM:signif}).

We conclude that for the expected shock parameters, $\epse \sim 0.3$ 
and $\epsB \simlt 10^{-3}$, the blast wave running into the ISM inevitably produces 
bright TeV emission.
A large fraction of observed bursts should have 
already been detected above $100$~GeV,
even if the prompt radiative efficiency 
exceeds 50\%, reducing the blast wave energy. The absence of detections disfavors
the ISM as the surrounding environment of most GRBs.

\subsection{TeV predictions for a sample of bursts with measured redshifts}

As discussed above,
the efficiency of TeV production and attenuation on the way to the observer
depends critically on the burst energetics and cosmological redshift.
For most bursts the redshifts and therefore also their energetics are unknown,
and one must resort to a statistical approach using
models for the luminosity function and redshift distribution.
On the other hand, for bursts that do have measured redshifts,
the uncertainties associated with the statistical approach can be avoided,
even though they are not sufficiently numerous to yield statistically reliable predictions 
for the entire GRB population.

\begin{figure} 
  \begin{center}
  \includegraphics[width=0.47\textwidth, trim=0 0 0 1cm]{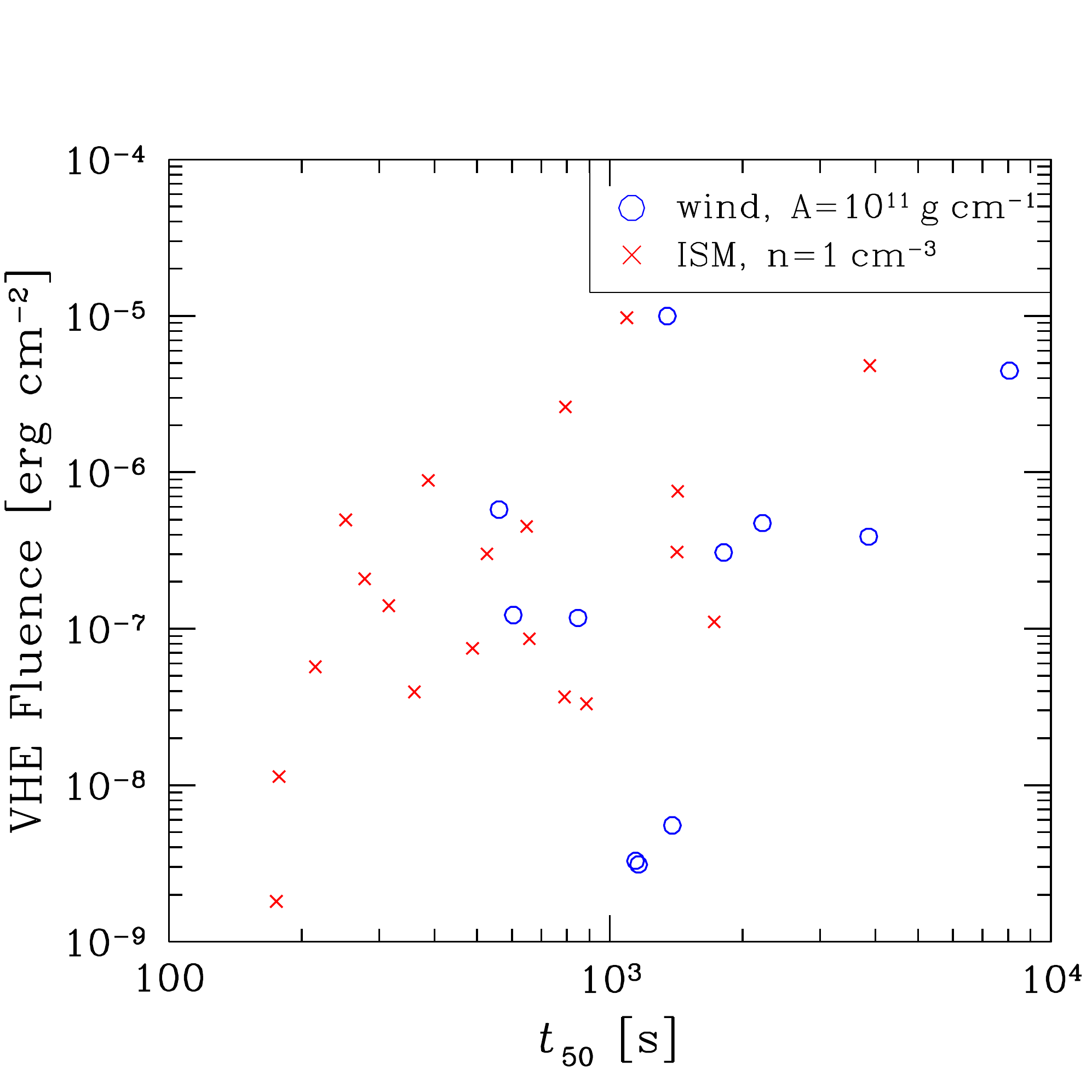}
  \vspace*{-0.4cm}
  \end{center}
  \caption{Predicted TeV properties of bursts with measured redshifts, for a sample compiled by Gruber et al. (2011).
  Blue open circles and red crosses correspond to 
  wind and ISM media, respectively.
  Parameters: radiative efficiency $\epsrad=0.5$, jet Lorentz factor $\Gjet=100$.
  }
  \label{Fig:Gruber:Fl_tobs}
\end{figure}

\begin{figure} 
  \begin{center}
  \includegraphics[width=0.47\textwidth, trim=0 0 0 1cm]{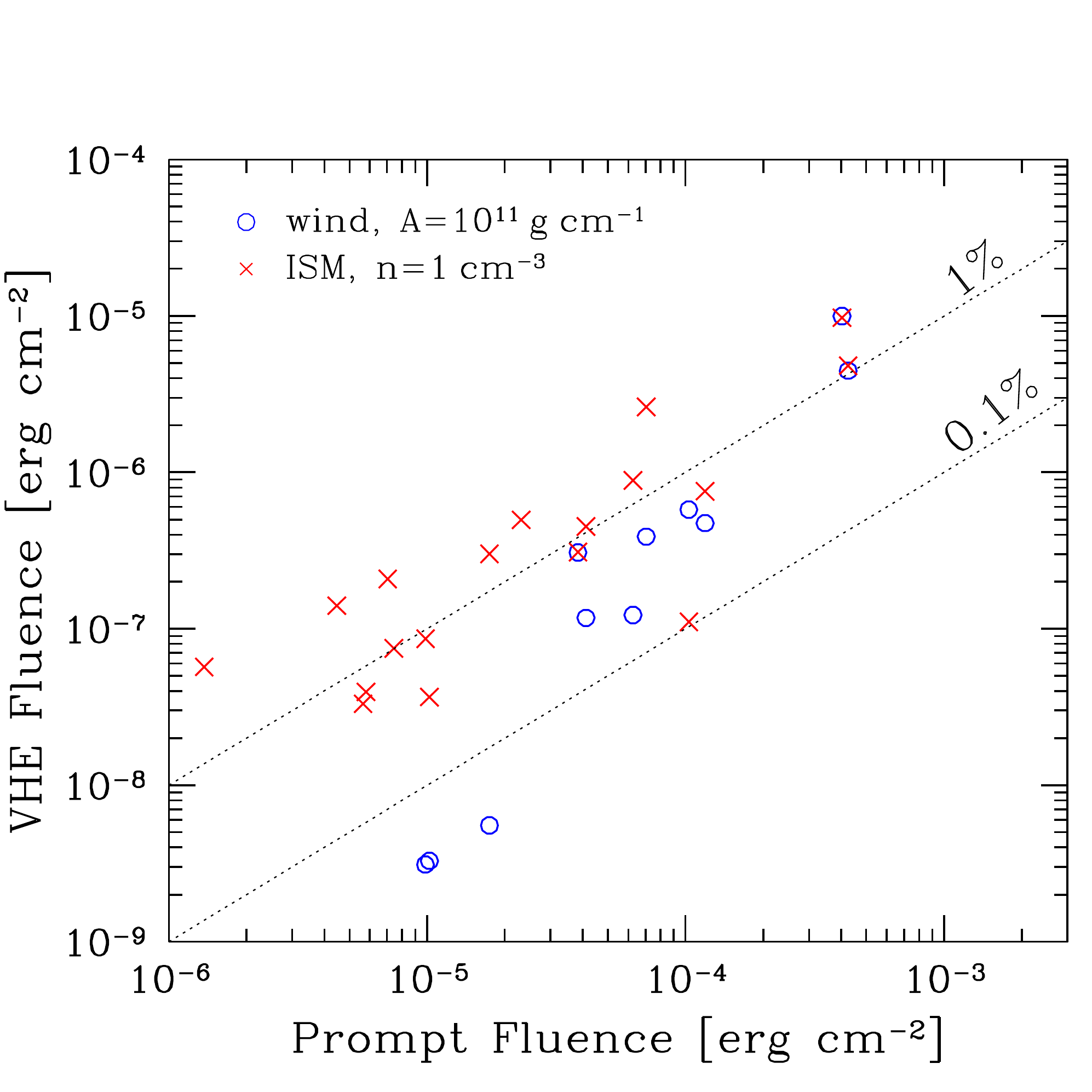}
  \vspace*{-0.4cm}
  \end{center}
  \caption{Same bursts as in Figure \ref{Fig:Gruber:Fl_tobs}, according to their measured prompt and predicted TeV fluences.
  }
  \label{Fig:Gruber:Fl_Fl}
\end{figure}

Figures \ref{Fig:Gruber:Fl_tobs} and \ref{Fig:Gruber:Fl_Fl} show
our predictions for TeV emission for a sample of GRBs with measured redshifts, compiled by
\citet{Gruber11}.
Out of the 30 bursts in the sample, 10 have redshifts $z>2$ and are strongly absorbed in 
the TeV band by the EBL; we excluded them from the sample also because
the EBL extinction models by \citet{Dominguez2011} used here only go up to $z=2$.
From the rest of the sample, the number of bursts detectable by CTA is
8 and 18 for a wind medium ($A=10^{11}$~g~cm$^{-1}$) and
ISM ($n=1$~cm$^{-3}$), respectively.

The distribution of the bursts over the reconstructed
TeV fluences and characteristic peak times
is consistent with 
the Monte-Carlo simulations presented in
Figures \ref{Fig:Fl_tobs}, \ref{Fig:Fl_Fl}, \ref{Fig:ISM:Fl_tobs}, \ref{Fig:ISM:Fl_Fl}.
In particular, it shows that the TeV emission from GRBs in wind media have a larger 
delay and significant suppression by $\gamma\gamma$ absorption.

\bigskip

\bigskip

\subsection{Radiative losses at the pair-dominated stage}

\label{sec:radloss}

At early stages of the blast wave expansion
the $\epm$ pairs created ahead of the forward shock by the prompt radiation
are sufficiently numerous to dominate the inertia of the
upstream plasma.
This occurs at $\xi\gtrsim 200$, when $\Zpm\gtrsim \mu\mprot/\me \approx 3000$ (B14).
The $\epm$ pairs then receive the bulk of the shock-generated heat
and promptly cool via IC emission,
making the forward shock almost 100\% radiatively efficient.
The prompt photons scattered by the medium ahead of the blast wave also penetrate into the unshocked jet
and generate $\epm$ pairs there; as a result, the radiative efficiency of the reverse shock is increased as well,
even though the pairs never come to dominate the inertia of the jet.

The losses of the blast-wave energy at
the pair-dominated stage depend on
two main factors:
\begin{enumerate}
\item
The fraction of the jet energy that is deposited into the blast wave during the radiatively efficient stage.
This fraction is less than unity if the reverse shock has not yet crossed the ejecta when the pair loading drops below $\Zpm\approx 3000$.
\item
The efficiency of dissipating the blast wave energy at
the radiative stage via the forward and reverse shocks.
This mainly depends on whether the reverse shock is relativistic;
a significant fraction of the jet energy
processed through the reverse shock
is dissipated in the relativistic regime.
Conversely, in the Newtonian regime the RS passes
through the ejecta without significant dissipation;
the dissipated energy in the forward shock is also small until the blast wave 
approaches $\Rdec$.
\end{enumerate}

The radiative regime ends at
\begin{align}
R_{\rm rad} \approx \frac{\sigmaT \LGRB}{8\pi\me c^3 \Gfs^2\xi} = 5.4\times 10^{15} \, \frac{L_{{\rm GRB},52}}{\Gamma_{{\rm fs},2}^{2}} \, \left(\frac{\xi}{200}\right)^{-1} \, \mbox{cm},
\label{eq:Rrad}
\end{align}
where $\LGRB$ is the prompt luminosity, $\Gfs$ is the Lorentz factor of the forward shock, and we have used the definition $\xi = \sigmaT\LGRB\tfs/(4\pi\me c^2 R^2)$,
where $\tfs=R/(2c\Gfs^2)$ is the time coordinate of the forward shock (B14).

The reverse shock reaches the end of the jet at
\begin{align}
\Rrs &= \frac{\Delta}{\bjet - \brs} \approx 2\Delta \left( \frac{1}{\Grs^2} - \frac{1}{\Gjet^2} \right)^{-1} \nonumber \\
&\approx 2c \tGRB \Grs^2 = 6\times 10^{15} \, t_{{\rm GRB},1} \, \Gamma_{{\rm rs},2}^2 \, \mbox{cm},
\label{eq:Rrs}
\end{align}
where $\bjet$ and $\brs$ are the velocities of the unshocked jet and reverse shock, respectively, and $\Gjet$ and $\Grs$ are the corresponding Lorentz factors.
In the second line of Equation (\ref{eq:Rrs}), the radial thickness of the ejecta has been approximated by $\Delta\approx c\tGRB$, and a relativistic reverse shock regime $\Gjet\gg\Grs$ has been assumed.

If both the forward and reverse shocks are radiative, the shocked plasma is
compressed into a thin shell, and one can
approximate $\Gfs\approx\Grs\equiv \Gamma$. 
Using this and taking the ratio of Equations (\ref{eq:Rrad}) and (\ref{eq:Rrs}),
one obtains the approximate fraction of jet energy
processed through the reverse shock by the end of the radiative stage
\begin{align}
\frac{R_{\rm rad}}{\Rrs} = 0.9 \, \frac{\E_{\rm GRB,53}}{t_{\rm GRB, 1} \Gamma^4_2} \left(\frac{\xi}{200}\right)^{-1}.
\label{eq:rradeff}
\end{align}
If $R_{\rm rad}/\Rrs \ge 1$, the blast wave is still radiative when the reverse shock reaches the end of the jet.
Note that the Lorentz factor of the plasma between the forward and reverse shocks cannot 
be arbitrarily high regardless of $\Gjet$;
its maximum value is given by (B14)
\begin{align}
\Gamma_{\rm max} \approx \left[ \frac{\gamma_{\rm pre} \Lkin}{8\pi A c^3 (1 + \Zpm\me/\mue\mprot)}
\right]^{1/4},
\end{align}
where $\Lkin\approx \Ekin/\tGRB$ is the kinetic luminosity of the jet,
and $\gamma_{\rm pre}$ is the pre-acceleration Lorentz factor of the ambient matter 
ahead of the forward shock. Equation~(\ref{eq:rradeff}) now yields
\begin{align}
\frac{R_{\rm rad}}{\Rrs} \gtrsim 0.61 \, \frac{A_{11}}{t_{\rm GRB,1}} \, \frac{\EGRB}{\Ekin} \, \left(\frac{\xi}{200}\right)^{-1}\,
\frac{(1 + \Zpm\me/\mue\mprot)}{\gamma_{\rm pre}},
\end{align}
where the last factor is typically $\sim 0.2$ at the end of the radiative stage.

As long as all the heat generated by the shock is radiated, the energy left in the 
shocked and cooled plasma is simply its rest mass multiplied by $\Gamma$.
Thus the fraction of radiated energy is $1-\Gamma/\Gjet$, which is close to unity when the reverse shock is relativistic,
or if the blast wave is already in the self-similar deceleration regime at $R_{\rm rad}$.
Overall, the fraction of energy lost by the blast wave at
the radiative stage is
\begin{align}
\frac{\E_{\rm rad}}{\Ekin} \approx \left(1-\frac{\Gamma}{\Gjet}\right) \, \min\left(1,\frac{R_{\rm rad}}{\Rrs}\right),
\end{align}
where $\Gamma$ is the blast wave Lorentz factor at $R_{\rm rad}$.
The loss can be substantial for fast jets exploding into dense media.

The simulations presented in previous sections
parametrize the blast wave kinetic energy as a fraction $(1-\epsrad)$
of the total energy budget
{\it after} the early pair-dominated stage,
which is the relevant quantity for determining TeV emission.
This minimizes the number of parameters,
since using the {\it initial} jet kinetic energy
introduces additional dependences on the burst duration as well as the jet Lorentz factor.
Furthermore,
there is little chance of measuring the jet kinetic energy before
it starts interacting with the surrounding medium.

\subsection{Instrument detection rates and comparison with earlier predictions}

The rates in Figures \ref{Fig:CTA:signif}, \ref{Fig:VER:signif} and \ref{Fig:VER:ISM:signif}
give the probability for detecting a burst assuming that it was quickly observed by the 
Cherenkov telescope at favorable observing conditions, at sufficiently small zenith angles. 
The number of detections is reduced when one takes
into account the duty cycle of the instrument and
the decreasing sensitivity (and the increasing low-energy threshold)
at zenith angles exceeding a few tens of degrees. Taking the duty cycle to be $10\%$,
one finds that only a few percent of the prompt triggers can be followed up by 
Cherenkov telescopes sensitive to photons of energies $\sim 100$~GeV
\citep[e.g.][]{Inoue2013}.

When our predictions are adjusted by this factor, we find that for {\it Swift}-like alerts 
($\sim 100$~yr$^{-1}$) the expected detection rate for CTA
is $\sim 0.2-0.3$~yr$^{-1}$ for a wind medium with $A=10^{11}$~g~cm$^{-2}$,
and $0.5-1$~yr$^{-1}$ for the ISM with $n=1$~cm$^{-3}$. For comparison,
{\it Fermi}/GBM provides $\sim 250$ alerts per year, however the localization 
error circle is typically larger than the LST field of view ($4.6\degree$ diameter).
A scanning mode operation could compensate for this, at the expense of the detected 
count rate \citep{Finnegan2011}. The analysis by \citet{Inoue2013} suggests
that the scanning mode yields comparable detection rates from GBM and {\it Swift} alerts.

The above rates can be compared with those obtained in earlier works using different 
methods. \citet{Gilmore2013} (see also \citealt{Gilmore2010a,Inoue2013})
used a phenomenological approach, approximating the high-energy spectrum as a 
power law with energy spectral index $\beta=1$,
normalized to yield a fixed LAT to GBM fluence ratio of $0.1$ (``fixed" model).
The redshift and prompt fluence distributions were
deduced from {\it Swift} and BATSE observations, respectively,
and were assumed to be uncorrelated with each other.
The ``detection efficiency" of \citet{Gilmore2013}
is analogous to the detection probability in Figure \ref{Fig:CTA:signif} for $S=5$ 
(i.e. $5\sigma$) and varies between $0.1-0.4$ for CTA,
under different assumptions about the EBL attenuation and CTA performance.
The predicted yearly detection rate was $\sim 0.5 - 1.5$~yr$^{-1}$.

\citet{Kakuwa2012} (se also \citealt{Inoue2013})
used the formation rate and luminosity function of GRBs by \citet{Wanderman2010},
and a phenomenological parametrization of the VHE spectrum similar to
\citet{Gilmore2013}. The predicted detection rates by CTA in the afterglow phase
were $0.13$~yr$^{-1}$ and $0.37$~yr$^{-1}$
for {\it Fermi}/GBM and {\it SVOM}/ECLAIRs
\citep{Paul2011} alerts, respectively.
Note that while EBL absorption was taken into account in the above papers,
intrinsic absorption by the X-ray afterglow was neglected.

The above papers focused mostly on the observational and instrumental aspects 
of the problem. By contrast, in this paper we focused more on the efficiency of VHE 
emission by GRB blast waves and its implications for the detection rates.
These complementary approaches help to make realistic
predictions for the detection rates.

\subsection{Caveats of the simplified IC model} 

\label{sec:caveats}

Our simplified model for high-energy emission only considers IC radiation from thermal electrons.
Furthermore, we did not explicitly calculate synchrotron emission
from either thermal or non-thermal particles;
instead, we used statistical information from observed X-ray afterglows
to randomly draw the energy and spectral slope of the afterglow spectrum
to find the target photon field for IC scattering as well as $\gamma\gamma$ absorption.
This was mainly done to (1) minimize the computation time of the high-energy spectra
and be able to make statistical predictions for the detectability of a large number of 
simulated bursts, and (2) to keep the number of parameters to a minimum,
in particular to avoid explicitly considering the non-thermal electron population.
Two caveats are associated with this approach:
\begin{enumerate}
\item
The X-ray afterglow properties only depend on the prompt energetics, and
are assumed to be statistically independent
of other parameters such as e.g. the external medium density, profile etc.
This is clearly not accurate for individual bursts.
However, one can reasonably expect our results to be approximately correct in a 
statistical sense, unless some unidentified correlation between certain parameters
systematically biases the expectation for the high-energy emission.
\item
The spectrum of the X-ray afterglow is assumed to be a simple power-law.
This disregards all the breaks in the spectrum associated with e.g. the fast/slow cooling transition, thermal synchrotron frequency
(akin to $\nu_{\rm m}$ corresponding to the low-energy end of the non-thermal electron distribution used in afterglow literature) etc.,
which would have to be calculated self-consistently for the electron distribution injected at the shock, downstream magnetization etc.
Again, one can hope that statistically these inaccuracies average out without systematically affecting our results.
\end{enumerate}


\section{Conclusions}

\label{sec:concl}

GRB explosions are expected to generate TeV emission
during the first minutes to hours after the explosion. The VHE radiation is efficiently
produced by IC upscattering 
of the X-ray afterglow photons by {\it thermal} electrons heated at the forward shock,
and can carry up to a few tens of percent of the explosion energy.
The emission can be significantly delayed with respect to the prompt MeV radiation, due to
(1) pair loading of the external medium by the prompt emission, which decreases the energy available per electron, and/or
(2) opacity seen by the TeV gamma-rays due to their collisions with the X-ray afterglow 
photons. 

The VHE emission must be bright until the characteristic IC energy of the upscattered 
photons in the forward shock drops below $\sim 100$~GeV.
This occurs when the blast wave is decelerated to Lorentz factor $\Gamma\approx 20$.
The emission mechanism is robust, since it only relies on the efficient 
heating of electrons in the forward shock ($\epse\sim 0.3$), which is seen in 
{\it ab initio} plasma simulations \citep{Spitkovsky2008a,SironiSpitkovsky2011}.
The same thermal mechanism of high-energy emission is
supported by the fitting of the GeV(+optical) flashes in several GRBs \citep{Hascoet15}.

We have simulated a large number of GRBs exploding in a wind medium and ISM, 
and examined their expected VHE luminosities, in particular their detectability with
current (VERITAS) and  upcoming (CTA) Cherenkov telescopes.
The cosmological redshifts, energetics, and afterglow properties of the simulated bursts were
drawn from the corresponding probability distributions inferred from observations.
Our simulations demonstrate that GRBs exploding into ISM must produce much 
brighter TeV emission compared with GRBs exploding into a wind of a 
Wolf-Rayet progenitor. This is in contrast to IC emission in the 0.1-10~GeV band, 
which is brighter for explosions in wind media and responsible for the early bright peak 
of the flash observed by Fermi LAT  (B14).

Our main conclusions may be summarized as follows:
\\
(1) The most striking result of the simulations is that low density ISM environments 
($n\sim 1-10$~cm$^{-1}$) are strongly disfavored by the lack of TeV detections with 
current Cherenkov telescopes. The high Lorentz factor and the large deceleration radius 
of the ISM blast wave makes it transparent to $0.1 - 1$~TeV photons. 
The peak of TeV emission occurs near the deceleration radius, when 
the jet is transferring its kinetic energy to the external blast wave. We find that the IC 
electron cooling (and hence TeV emission) is efficient at this stage, and a significant 
fraction of the explosion energy (comparable to $\epse$) is radiated in the TeV band.
This luminous VHE emission is hard to avoid in the ISM scenario and should 
have already been detected by current instruments.
\\
(2) From 
the fact that none of the $>50$ rapid follow-up observations by 
Cherenkov telescopes have resulted in a detection, we conclude that the majority of GRBs 
exploded into dense environments,
consistent with a Wolf-Rayet wind. This conclusion is also consistent with the
reconstruction of the early GeV(+optical) emission observed by \Fermi LAT
(B14; \citealt{Vurm14,Hascoet15}). 
For a very dense wind, $A=\rho R^2\gg 10^{11}$~g~cm$^{-3}$, the blast wave
decelerates so early that its thermal
TeV emission ceases before the source becomes transparent to VHE gamma-rays.
Then TeV detection is not expected. 
The transition to this regime occurs around $A\sim 3\times 10^{11}$~g~cm$^{-1}$.
\\
(3) The wind density parameter $A$ reconstructed from the GeV+optical data for 7 bright
bursts \citep{Hascoet15} is comparable to $10^{11}$~g~cm$^{-1}$. Therefore, $A$ is 
not expected to be much above $10^{11}$~g~cm$^{-1}$ for bright GRBs, and hence a 
large number of bursts should produce TeV emission detectable with CTA. Our detailed 
Monte-Carlo simulations predict a high detection rate for $A=10^{11}$~g~cm$^{-1}$
and weaker but still detectable VHE emission for $A$ up to 
$5\times 10^{11}$~g~cm$^{-1}$ (Figures~\ref{Fig:Fl_tobs} and \ref{Fig:Fl_Fl}).

It is well known that $\gamma\gamma$-absorption on the extragalactic background 
light (EBL) strongly attenuates TeV emission from cosmological sources.
Only a few percent of GRBs which are most nearby could be detectable above 
$200-300$~GeV. At average GRB redshifts, the EBL attenuation becomes significant 
already around $100$~GeV, which underscores the importance of continued efforts to 
lower the photon energy threshold of Cherenkov instruments. 
Such efforts will almost certainly be rewarded since we know from {\it Fermi}/LAT 
observations that GRBs do emit photons above a few tens of GeV.

\acknowledgments
We thank Jochen Greiner for providing the data for a sample of bursts with measured 
redshifts, and Brian Humensky for discussions of the Cherenkov telescopes.
This work was supported by NSF grant AST-1412485, NASA grant NNX15AE26G,
and a grant from the Simons Foundation (\#446228, Andrei Beloborodov).

\bibliographystyle{apj} 
\bibliography{biblio}

\end{document}